\documentclass[aps,prb,preprint,12pt,nofootinbib]{revtex4-1}  %Required for AJP (divide by 3 for # AJP 2-col pages)
\usepackage{amsmath}  % needed for \tfrac, \bmatrix, etc.
\usepackage{amsfonts} % needed for bold Greek, Fraktur, and blackboard bold
\usepackage{graphicx} % needed for figures
\usepackage{color}

\newcommand{\be}[1]{\begin{equation}\label{#1}}
\newcommand{\ee}{\end{equation}}
\newcommand{\bea}[1]{\begin{eqnarray}\label{#1}}
\newcommand{\eea}{\end{eqnarray}}
\newcommand{\no}{\nonumber \\}
\newcommand{\Fig}[1]{Fig.(\ref{#1})}

\newcommand{\Eq}[1]{Eq.(\ref{#1})}

\newcommand{\Sec}[1]{Section~\ref{#1}}
\newcommand{\bsub}{\begin{subequations}}
\newcommand{\esub}{\end{subequations}}
\newcommand{\bwt}{\begin{widetext}}
\newcommand{\ewt}{\end{widetext}}

\usepackage{color}

\def\black#1{{\color{black} {#1}}}
\def\trm#1{\textrm{#1}}
\def\tit#1{\textit{#1}}

\def\ttt#1{\texttt{#1}}
\newcommand{\om}{\omega}
\newcommand{\Om}{\Omega}

\def\a0{{\alpha_0}}

\def\da0{{\dot{\alpha}_0}}

\def\myoverDefn#1#2{\hbox{\space \raise-2mm\hbox{$\textstyle{#1} \atop \scriptstyle{#2}$} }}
\def\defn{\overset{\textrm{def}}{=}}
\def\om{{\omega}}

\def\t{{\tau}}

\def\mA{{\mathcal{A}}}

\def\mO{{\mathcal{O}}}

\def\G{{\Gamma}}

\def\g{{\gamma}}
\def\d{{\delta}}
\def\a{{\alpha}}

\def\rp{r_{P}}
\def\rp2{r_{p}^{2}}

\def\tX{\tilde{X}}

\newcommand{\half}{\frac{1}{2}}

\newcommand{\pd}{\partial}

\def\eps{\epsilon}
\def\a0{{a_0}}
\def\b_0{{\beta_0}}
\def\b{{\beta}}
\def\tT{\tilde{T}}
\def\tX{\tilde{X}}
\begin{document}
%====================================
% Be sure to use the \title, \author, \affiliation, and \abstract macros
% to format your title page.  Don't use lower-level macros to  manually
% adjust the fonts and centering.

\title{Inertial-to-Rindler Coordinates, with applications to the Twin Paradox, Radar Time and the Unruh Temperature}
% In a long title you can use \\ to force a line break at a certain location.

%When submitting the manuscript for review, do not include the author's name or institution
%\author{Daniel V. Schroeder}
%\email{dschroeder@weber.edu} % optional
%\altaffiliation[permanent address: ]{101 Main Street, Anytown, USA} % optional second address
% If there were a second author at the same address, we would put another 
% \author{} statement here.  Don't combine multiple authors in a single
% \author statement.
%\affiliation{Department of Physics, Weber State University, Ogden, UT 84408-2508}
% Please provide a full mailing address here.

%\author{David P. Jackson}
%\email{ajp@dickinson.edu}
%\affiliation{Department of Physics, Dickinson College, Carlisle, PA 17013}
\author{Paul M. Alsing}
\thanks{Extended version. Incorporates pedagogical peer-review edits accepted in Am. J. Phys. (under the title \tit{Inertial-to-Rindler Coordinates, with applications to the Twin Paradox, Radar Time, and Considerations for an Adiabatic Unruh Temperature}, scheduled to be published in March 2027), alongside expanded technical derivations in \black{Sec. V} omitted from the print version, plus included code in the Appendix\hspace{-0.05in}
%Standalone technical follow-up paper is cross-referenced.
}
\email{alsingpm@gmail.com} % optional
\altaffiliation[permanent address: ]{Florida Atlantic University, 777 Glades Rd, Boca Raton, FL, 33431, USA} 
% See the REVTeX documentation for more examples of author and affiliation lists.

\date{\today}

\begin{abstract}
In this work we formulate a two-parameter family of transformations in flat Minkowksi spacetime that smoothly interpolates between motion with constant initial/final velocity (inertial coordinates), and with constant acceleration (Rindler coordinates \cite{Rindler:1960,Rindler:1966}), which we term Inertial-to-Rindler (I2R) coordinates. 
We revisit the Twin ``Paradox" and show how the new I2R coordinates justify the ``immediate-" and ``gradual-turnaround" scenarios discussed in many texbooks and articles.
We also examine the radar time formulation of hypersurfaces of simultaneity by Dolby and Gull \cite{Dolby_Gull:2001} for these new coordinates as we transition from zero to uniform acceleration.
Finaly we re-examine the negative frequency content of a  purely positive frequency Minkowski plane wave as observed by the I2R observer, and derive perturbative corrections to the Unruh \cite{Unruh:1976} temperature for the two cases of initial/final velocities slightly greater than zero, and slightly less than the speed of light - the latter of which characterizes constant acceleration motion. 
We argue for a proposed velocity-dependent generalization of the Unruh temperature that smoothly varies from zero at zero-acceleration, to the standard form at constant acceleration.
\end{abstract}
% AJP requires an abstract for all regular article submissions.
% Abstracts are optional for submissions to the "Notes and Discussions" section.

\maketitle % title page is now complete

%=========================================================
\section{Introduction}\label{sec:Intro}
%=========================================================
Inertial observers moving at constant velocity relative to one another form the basis of Einstein's 
Special Relativity (SR), along with the constancy of the speed of light. Within SR, and then when moving to General Relativity (GR), one learns of the central importance of an observer moving with constant acceleration \cite{Rindler:1969,MTW:1970,DInverno:1992,Hartle:2003,Ryder:2009} described by the celebrated Rindler coordinates \cite{Rindler:1960,Rindler:1966}. Hawking's seminal work showing that black holes  (BH) evaporate \cite{Hawking:1975}, and hence are  classical thermodynamic objects, relied on a stationary observer hovering just outside the horizon of a BH at constant radius, and hence undergoing constant acceleration in order to maintain stationary. The Hawking temperature of a BH of mass $M$ is proportional to its surface gravity, which is the gravitational acceleration $\kappa = G M/R_s^2$ at \black{the} Schwarzschild radius $R_s = 2 G M/c^2$ of its horizon. 
One year later, Unruh \cite{Unruh:1976} showed that essentially the same effect can occur in flat (Minkowski) spacetime for non-inertial observers undergoing constant acceleration $a_0$. Analogously, the Unruh temperature is proportional to the (Rindler) observer's acceleration. These two scenarios are intimately related, since in GR we learn that the metric of the spacetime for a stationary observer just outside the BH can be put into the form of the constant acceleration Rindler metric \cite{Rindler:1969,MTW:1970,DInverno:1992,Hartle:2003,Ryder:2009}. 

The Rindler coordinates for an object located at  $(t,x)$ in the frame of the accelerated observer
 (Rob for short, with origin $x=0$) undergoing constant acceleration $a_0$ with respect to an inertial Minkowski observer (Bob, for short) with coordinates $(T, X)$ are given by the well-known transformation \cite{Rindler:1960,Rindler:1966}
\be{Rindler:coords:t:x}
T = \frac{1}{a_0}\, e^{a_0 x}\,\sinh(a_0 t), \qquad 
X = \frac{1}{a_0}\, e^{a_0 x}\,\cosh(a_0 t),
\ee
where we have taken the speed of light to be $c=1$ \black{(which we will use throughout this work).
\black{Dimensional} units can be obtained by reinserting appropriate factors of $c$ (e.g. $a_0\,x \to a_0\,x/c^2$, $a_0\,t \to a_0\,t/c$, $T\propto 1/a_0\to c/a_0$ and  $X\propto 1/a_0\to c^2/a_0$
in \Eq{Rindler:coords:t:x} above, etc\ldots)}. The velocity of the accelerated observer \black{(which by definition is a massive object)} with respect to the inertial observer is given by $-1\le V(t) = dX(t)/dT(t) = \tanh(a_0 t) \le 1$. Thus, the uniformly accelerated observer \black{apparently} begins their journey \black{asymptotically in time} at the speed of light $V(-\infty)~=~-1$, traveling with decreasing velocity towards the inertial observer's origin $X~=~0$, slows down to $V(0)=0$ there, and then continues back out to infinity, again reaching a terminal velocity of the speed of light $V(\infty) = 1$. 
%Conceptually therefore, 
\black{At first glance, there appears to be} 
%there is 
a technical inconsistency which  never causes any real trouble in physical calculations, but is nonetheless theoretically bothersome: namely, the physical (massive) observer can never achieve the speed of light $c=1$ which the uniform accelerated observer begins and ends their journey. 
\black{However, the massive observer only asymptotically approaches the speed of light at $T\to\pm\infty$, so in actuality, there is no real physical inconsistency. Nonetheless, a}
%A 
much more satisfying situation would be one in which Rob begins and ends with a finite speed $\pm \beta_0$ where $0<\beta_0 <1$, while still undergoing some manner of, now non-uniform, acceleration $a(t)$. This is the core subject of investigation of this paper, which we develop shortly. We term these new coordinates Inertial-to-Rindler coordinates, and denote them as I2R coordinates. While the Rindler coordinates are a 1-parameter set of transformations between $(T,X)$ and $(t,x)$ with acceleration parameter $a_0$ (since the terminal velocity is taken to be fixed at 
$\beta_0=1$), the new I2R coordinates involve  two parameters $(a_0, \beta_0)$, which one can choose freely, and reduce (for $x=0$) to constant velocity and uniform acceleration as 
$\beta_0\to 0$ and $\beta_0\to 1$, respectively, for constant $a_0$, namely,
\bea{}
T &\to& t, 
\hspace{1.0in} X \to \beta_0\,t  = \beta_0\,\black{T}, 
\hspace{0.25in} \trm{for}\quad \beta_0\ll 1,
\hspace{0.15in} \trm{constant velocity}, \\
T &\to& \frac{1}{a_0}\,\sinh(a_0 t), \quad 
X \to \frac{1}{a_0}\,\cosh(a_0 t), 
\quad \trm{for}\quad \beta_0\to 1,\quad \trm{constant acceleration}.
\eea{} 
We shall show below that the proper acceleration $a(t)$ also depends on $(a_0, \beta_0)$ and is approximately constant at its peak value $a(0) = \beta_0 a_0$ for an adjustable finite duration $(-t^*, t^*)$, symmetric about the origin, and essentially zero outside this range. By adjusting both the value of $\beta_0$ and $a_0$ one can smoothly (continuously) create either an ``immediate" turnaround, or a ``gradual" turnaround scenario for the twin ``paradox," which is usually handled by a piecewise approximation \cite{Dolby_Gull:2001} of (i) an inbound (inertial) initial velocity of $-\beta_0$ from $t\in(-\infty,t^*)$, (ii) (Rindler) uniform acceleration for a brief time 
$(-t^*, t^*)$, and (iii)~returning with (inertial) outbound velocity of $\beta_0$ from $(t^*, \infty)$. Inspired by the work of Dolby and Gull \cite{Dolby_Gull:2001}, we also explore the coordinate free \tit{radar time} assignment of consistent lines of simultaneity for  the I2R observer, which we designate as Irwin, for short. As shown by Dolby and Gull, the radar time assigns a \tit{unique} time to any event (to which the traveling twin can send and receive signals), resolving misconceptions that have appeared in the literature and textbooks claiming the lines of simultaneity can ``cross" when an ``immediate" turnaround traveler's velocity abruptly ``swings around" from 
$-\beta_0$ to $\beta_0$ (thus inconsistently having them assign   multiple different coordinates to the same event). 

Turning to how the traveling observer perceives an inertial observer's plane wave, we know that 
for inertial observers undergoing constant velocity $\beta_0$, Bob's (rest frame) positive frequency $\om$  plane waves  remain positive, Doppler shifted, frequency plane waves with frequency
$\om' = \sqrt{\tfrac{1\pm \beta_0}{1\mp \beta_0}}\,\om$ (where the upper/lower signs corresponds to blueshifts and redshifts, respectively) as perceived by the moving observer.
However, the uniformly accelerating  observer Rob perceives the purely positive frequency plane wave of Bob (inertial) as having both positive and negative frequency components when Fourier analyzed. 
The absolute square of the negative frequency component has a (boson) thermal Planck spectrum with an Unruh temperature proportional to the constant acceleration $a_0$, and is an indication of the particle creation that Rob perceives as he travels through  the surrounding inertial vacuum. Ultimately, this particle creation results from the constant force Rob undergoes to keep him in uniform acceleration. In fact, any motion that is not zero acceleration will produce a response in Rob's particle detector \cite{Birrell_Davies:1982}, though the exact thermal nature of the Unruh radiation for uniform acceleration is directly tied to the hyperbolic nature of the Rindler transformation of coordinates \cite{Alsing_Milonni:2004,Padmanabhan:2010}. Rob's orbit is confined to the Right Rindler Wedge (RRW) of Minkowski space $X>|T|>0$ which is causally disconnected from any ``mirror" observer 
\black{(who we name)} ``boR" traveling in the Left Rindler Wedge (LRW)
$X<-|T|<0$. 
\black{
This is easily seen from \Eq{Rindler:coords:t:x} where $X/T\overset{t\to\pm\infty}{\longrightarrow}\pm 1$ independent of all Rindler spatial locations $\xi$. Equivalently, the velocity profile of the Rindler observer is by definition given  as $\beta(t) \defn dX(t)/dT(t) = \tanh(a_0\,t)$, so that 
$\beta(t)\overset{t\to\pm\infty}{\longrightarrow}\pm 1$ are the asymptotes of the Rinder observer's orbits for any fixed coordinate $\xi\defn\frac{1}{a_0} \ln(1+a_0 x)$ (defined after \Eq{Rindler:metric}).
} % end of red
Quantum mechanically, the inertial vacuum is an entangled 
(two-mode squeezed \cite{Scully_Zubairy:1997,Agarwal:2013,Gerry_Knight:2023}) state of the LRW and RRW. Since Rob's orbit is confined soley to the RRW and can never receive any signals from 
boR in the LRW, Rob's quantum mechanical state is formed by tracing over the states of the LRW \cite{Unruh:1976,Birrell_Davies:1982,Carroll:2004}, resulting in a thermal density matrix, with the Unruh temperature
$k_b T_U \defn \tfrac{\hbar\, a_0}{2 \pi c}$, where $k_b$ is Boltzmann's constant. This temperature could argued from dimensional grounds alone by noting that $k_b T_U\sim \hbar\,\om_0$ for some frequency $\om_0$. The only frequency in the problem is $\om_0 = a_0/c$ in flat spacetime. In curved spacetime, the relevant frequency in the problem is the surface gravity $\kappa = G M/R^2_s = c^4/(4 G M)$ where $R_s = 2 G M/c^2$ is the Schwarzschild radius for a body of mass $M$. Thus, the Hawking temperature of a radiating BH follows from the Unruh temperature by (essentially) setting $a_0\to\kappa$, i.e. 
$k_b T_U \defn \tfrac{\hbar\, \kappa}{2 \pi c} = \tfrac{1}{8 \pi}\,\left(\tfrac{M_P}{M}\right)\, M_p\,c^2$, where
$M_P = \sqrt{\tfrac{\hbar\,c}{G}}$ is the Planck mass. Again, the Hawking temperature results from the \black{observation} of a stationary observer sitting outside the BH at fixed radius $R>R_s$, thus undergoing uniform acceleration - hence it's connection to the Unruh effect.

In the following we will re-examine the Unruh temperature for the observer Irwin undergoing an I2R-trajectory that  (i) lies within the RRW \black{(of Rob)} for a duration when his acceleration is effectively non-zero, and (ii) essentially constant velocity outside the RRW. 
We propose and argue that one can define an I2R temperature given by 
$k_b T_{I2R} \defn \tfrac{\hbar\, a(t)}{2 \pi c}$ where $a(0) = \beta_0\,a_0$, analogous to the Unruh temperature $T_U$ with essentially $a_0\to \beta_0\,a_0$ during the period that Irwin is in the RRW undergoing non-zero acceleration, which is approximately constant at $a(0)$.
$T_{I2R}$ has the correct limits of (i) zero temperature for constant velocity, i.e. $a_0=0$, or nearly zero for very short periods of acceleration with $\beta_0\to 0$ for fixed $a_0$, and (ii) the Unruh temperature for uniform acceleration 
$\beta_0\to 1$ with $a(0)= \b_0\,a_0\to a_0$.
\black{Note that in this work, whenever we refer to the terms 
RRW ($X\ge |T|)$, or LRW ($X\le0, |X|>|T|)$, we always mean those regions associated with the uniform accelerated trajectory $a_0$ of Rob. }

The outline of this paper is a follows.
In \Sec{sec:Rindler} we review the derivation of the Rindler coordinates, which we later extend to the derivation of the I2R coordinates. We also briefly review the Rindler metric and Unruh effect for uniform acceleration, which will subsequently be generalized to I2R coordinates.
In \Sec{sec:I2R} we derive the I2R coordinates and examine its proper acceleration for an observer Irwin located at his origin, and also derive the I2R metric. We re-examine the twin ``paradox" and consider the inertial time $T(t)$ as a continuous (vs a conventional piecewise) function of Irwin's proper time $t$, as a function of $0\le \beta_0\le 1$, for a fixed value of the parameter $a_0$. 
In Section IV we derive the radar time for lines of simultaneity for Irwin in Bob's inertial coordinates  
$(T(t),X(t))$ as a function of \black{Irwin's} proper time $t$. We explore cases for 
$\beta_0 = \{0.0001, 0.1, 0.25, 0.5, 0.75, 0.9999\}$.
In Section V we re-examine the Unruh temperature by considering the  spectrum of both the positive and  negative frequency content of  Bob's inertial, purely positive frequency, plane waves as perceived (Fourier analyzed) by Irwin with respect to the his proper time $t$. The resulting integrals cannot be done in closed form (except of course for the extreme limits $\beta_0=0$ and $\beta_0=1$), so we approximate them for the two cases 
(i) $\beta_0 = \eps \ll 1$, and (ii) $\beta_0 = 1-\eps \lesssim 1$ to first order in $\eps$. 
For the former case (i) we obtain Doppler shifted positive frequencies and no negative frequency components.
For the latter case (ii) we compute  first order corrections to the Unruh temperature.
In general, we argue for why an I2R temperature defined by 
$k_b T_{I2R} \defn \tfrac{\hbar\, a(t) }{2 \pi c}$ makes sense for the period of Irwin's non-zero acceleration, essentially for the period of time while Irwin is in the RRW, and causally disconnected from the LRW. 
Finally, in  Section VI, we present our conclusions and prospects for future work.

%====================================================================
\section{Uniform Acceleration, Rindler coordinates and the \\Unruh effect}\label{sec:Rindler}
%=====================================================================
The most direct method for deriving the Rindler transformation \Eq{Rindler:coords} between 
the inertial coordinates $(T, X)$ for Bob, and the uniformly accelerated coordiantes $(t,x)$ for Rob, is to first consider Rob to be at his origin $x=0$ with coordinates $(T(t), X(t))$. 
% and to specify the desired velocity profile $dT/dX = \beta(t) = \tanh(\a0 t)$. 
Let us write Bob's inertial 4-velocity of Rob as 
$u^\mu = (T'(t), X'(t))\defn (\g(t), \g(t)\beta(t))$ where primes denote differentiation with respect to Rob's proper time $t$, and we have denoted $\g(t) \defn 1/\sqrt{1-\beta^2(t)}$.
Using the Minkowski metric with signature $ds^2 = dT^2 - dX^2$ we have quite generally
$(u^0)^2-(u^1)^2 = \g^2(t) -  \g^2(t)\beta^2(t)~\equiv~1$.

We now specify the desired velocity profile, which for the Rindler case of constant acceleration $a_0$ is 
$\beta(t) = \tanh(a_0 \,t)$. Thus, using the identity $1-\tanh^2(x) = 1/\cosh^2(x)$ we have the simple equations
$T'(t) = \g(t) = \cosh(a_0 t)$ and $X'(t) = \g(t)\,\beta(t) = \sinh(a_0 t)$ which trivially integrate to yield
\be{Rindler:coords} 
T(t) = \frac{1}{a_0}\, \sinh(a_0 t), \qquad  
X(t) = \frac{1}{a_0}\, \cosh(a_0 t).
\ee
It is easy to see that $a_0$ is Rob's proper acceleration by computing 
$a^\mu = a_0\, (\sinh(a_0 t), \cosh(a_0 t)) = a_0^2\,(T(t), X(t))$, and computing
$a(t) \defn \sqrt{a^\mu(t)\,a_\mu(t)} = a_0$. Rob's trajectory at his spatial coordinate origin $x=0$ is given by the hyperbola $X^2-T^2 = 1/a_0^2$, which are confined to the RRW quadrant $X>|T|>0$. Lines of (simultaneity) constant time $t_0$ are given by lines of constant slope through the origin 
$T = \tanh(a_0 t_0) X$, for $-\infty < t_0 < \infty$.

\subsection{Rindler coordinates $\mathbf{(t,x)}$ for uniform acceleration}
To find coordinates $(t,x)$ in the non-inertial Rindler frame, we follow the arguments of 
Padmanabhan \cite{Padmanabhan:2010}, and in particular, Gelis (p341)  \cite{Gelis:2021}. Consider a rigid rod of proper length $x$, with one end at Rob's origin and the other at fixed position $x$. The ends of the rod follow the worldlines in Rob's co-moving frame of
$r^\mu_{left} = (t,0)$ and $r^\mu_{right} = (t,x)$. These are also the coordinates of the endpoints of the rod in the instantaneous co-moving rest frame (since at any particular instant the rod is at rest in the co-moving frame). Define the vector $\Delta^\mu = r^\mu_{right}-r^\mu_{left} = (0,x)$ in the co-moving frame.
Since the co-moving frame (Rob's orgin) is inertial, a simple Lorentz transformation gives 
$\Delta^\mu$ in the rest frame (Bob)
\be{Gelis:p341}
\begin{pmatrix}
	 \Delta^0  \\
	 \Delta^1 \\
\end{pmatrix}_{\trm{Bob}}
=
\begin{pmatrix}
	 \g_0 & \beta_0\g_0 \\
	\beta_0\g_0 & \g_0 \\
\end{pmatrix}
\begin{pmatrix}
	0  \\
	 x \\
\end{pmatrix}
= 
\begin{pmatrix}
	 u^0 &u^1 \\
	u^1 & u^0 \\
\end{pmatrix}
\begin{pmatrix}
	0  \\
	 x \\
\end{pmatrix}
= 
\begin{pmatrix}
	 u^1 \, x \\
	u^0 \, x \\
\end{pmatrix}
=
\begin{pmatrix}
	 X'(t) \, x \\
	T'(t) \, x \\
\end{pmatrix}.
\ee
%=================
\black{Here, $\Delta^0$ and $\Delta^1$ are the temporal and spatial differences respectively, between the events that occur at $(0,0)$ and $(0, x)$, but viewed in the inertial frame moving in the negative $x$-direction with speed $\beta_0 c$.} %end red
%=================
In the rest frame (Bob) the endpoints of the rod are located at $(X^0(t), X^1(t))$ (which is the position of the accelerated observer Rob, holding one end of the rod at $x=0$), and at $(T,X)$ at the other end.
Therefore, we have
\bea{Gelis}
T - \frac{1}{a_0} \sinh(a_0 t) &=&   \sinh(a_0 t) \,x
\; \Rightarrow \; 
T = \frac{1 + a_0\,x}{a_0}\, \sinh(a_0 t)
\; \Leftrightarrow \; 
t = \frac{1}{2\,a_0} \ln\left(\frac{X+T}{X-T}\right), \qquad \label{Gelis:line1}
\\
X - \frac{1}{a_0} \cosh(a_0 t) &=&  \cosh(a_0 t)\, x 
\; \Rightarrow \; 
X = \frac{1 + a_0\,x}{a_0}\, \cosh(a_0 t)
\; \Leftrightarrow \; 
x = -\frac{1}{a_0} + \sqrt{X^2-T^2}. \qquad \label{Gelis:line2}
\eea
The metric is given by 
\be{Rindler:metric}
ds_R^2 = dT^2 - dX^2 = (1 + a_0\,x)^2\,dt^2 - dx^2.
\ee
Since $a_0$ is constant, we can define another spatial coordinate $\xi$ by $dx = (1+a_0\,x)\,d\xi$
or $(1+a_0\,x) = e^{a_0\,\xi}$, which allows us to put the Rindler transformation into the form
of \Eq{Rindler:coords:t:x}, with the conformally flat metric
\be{Rindler:metric:conformally:flat}
ds_R^2 = e^{2 a_0 \xi}\,(dt^2 - d\xi^2), \quad 
T = \frac{1}{a_0}\,e^{a_0 \xi}\, \sinh(a_0 t), \quad 
X = \frac{1}{a_0}\,e^{a_0 \xi}\, \cosh(a_0 t).
\ee
%
%Since the Rindler metric in this form is conformally flat, it admits 
%(unnormalized) positive frequency ($e^{-i \om t}$) and negative frequency ($e^{i \om t}$), right moving plane waves of the form $e^{i (k\,x \mp\,\om t)} = e^{i \om (x \mp  t)}$, where we have taken $\om = |k| >0$.
%This is analogous to the positive and negative frequency, right moving plane waves of the Bob's unaccelerated rest frame $e^{i \Om (X \mp T)}$. 
Since the Rindler metric in this form is conformally flat, it admits 
(unnormalized) positive frequency ($e^{-i \om t}$)
%(\black{conventionally defined as} $e^{-i \om t}$) 
and negative frequency ($e^{i \om t}$), 
right \black{($kx - \om t$),} 
\black{and left \black{($kx + \om t$)},}
moving   plane waves of the form $e^{i (k\,x \mp\,\om t)} = e^{i \om (x \mp  t)}$, where we have taken $\om = |k| >0$.
This is analogous to the positive and negative frequency, right moving plane waves of Bob's unaccelerated rest frame $e^{i \Om (X \mp T)}$.

\subsection{The Unruh effect}\label{subsec:Unruh:effect}
Let us parameterize a Lorentz transformation (LT) by the \tit{rapidity} parameter $r$ viz.
$\g_0 = \cosh r$ and $\beta_0\g_0 = \sinh r$ with $\beta_0 = \tanh r$. Then the LT between 
$\trm{Bob}'$ with coordinates $(\tT, \tX)$ moving with constant velocity $\beta_0$ with respect to the rest frame of Bob (with coordinates $(T,X)$) is given by
\be{LT}
T = \cosh r\, \tT + \sinh r\, \tX, \qquad X = \sinh r\, \tT + \cosh r\, \tX.
\ee
Bob's plane waves $e^{i \om (X \mp T)}$ is seen by $\trm{Bob}'$ by substituting in \Eq{LT} to obtain $e^{i \om' (\tX \mp \tT)}$ with $\om' = e^{\mp r}\,\om$. Therefore, $\trm{Bob}'$ perceives Bob's positive and negative frequency plane waves to remain positive and negative frequency, respectively, but Doppler shifted by a constant factor  $e^{\mp r} = \sqrt{\tfrac{1\mp\beta_0}{1\pm\beta_0}}$. 

We can now perform the same operation by substituting in the Rindler transformation \Eq{Rindler:metric:conformally:flat}
to obtain $e^{i \om (X \mp T)} \to e^{i \om'_R (x \mp t)}$ where the frequency $\om'_R$  perceived by Rob is now ``chirped," i.e. $\om'_R = (\om/a_0)\,e^{\mp a_0 t}$. Thus, modulo a rescaling of $\om$ by a factor of $1/a_0$, in essence the Doppler shift perceived by Rob is exponential in time, i.e. akin to taking $r\to a_0 \,t$.
This results, because at each instant of time, constant acceleration implies that the velocity 
used to Lorentz transform into the instantaneous co-moving frame for Rob increases by $\Delta \beta(t)$, so that the resulting Doppler shift is now time dependent.  

We now inquire as to the frequency content of Bob's positive/negative frequency plane waves 
$e^{\mp i \om U}$, where we have defined $U=T-X$ (and $V=T+X$ as \tit{lightcone} coordinates).
We decompose a right-moving purely positive frequency inertial plane wave $\tfrac{1}{\sqrt{\om}}\,e^{-i \om U}$ into the positive and negative right-moving plane waves of a Rindler observer $\tfrac{1}{\sqrt{\Om}}\,e^{-i \Om t}$ viz
\bea{alpha:beta}
\hspace{-0.5in}
\tfrac{1}{\sqrt{\om}}\,e^{-i \om U} = 
\tfrac{1}{\sqrt{\Om}}\,\int_{-\infty}^{\infty} d\Om\,
\left(
\alpha_{\om \Om}\,e^{-i \Om t} + \beta_{\om \Om}\,e^{i \Om t}
\right)
&\Rightarrow&
\begin{bmatrix}
\alpha_{\om \Om} \\
\beta^*_{\om \Om} \\
\end{bmatrix}
= \frac{1}{2 \pi}\,\sqrt{\frac{\Om}{\om}}\,
\int_{-\infty}^{\infty} dt\, e^{\mp i \om U}\, e^{i \Om t}, \no 
\hspace{-0.5in}
&\Rightarrow&
\begin{bmatrix}
|\alpha_{\om \Om}|^2 \\
|\beta_{\om \Om}|^2 \\
\end{bmatrix}
\propto
\left| 
\black{\mA_\mp\defn} % end red
\int_{-\infty}^{\infty} dt\, \ e^{i \Om t\, \mp\, i\,\left(\tfrac{\om}{a_0}\right)\,e^{-a_0 t }} 
\right|^2,\qquad
\eea
\black{where we have used \Eq{Rindler:coords} to evaluate $U=T-X = -\tfrac{1}{a_0}\,e^{-a_0 t}$ in the second line above,
}%end of red
and  the Fourier transform has been performed with respect to Rob's proper time $t$.
\black{$U$ is the coordinate for  a right moving wave since it is constant on lines 
at $45^\circ$
where as $T$ increases, $X$ increases.
Since in free space there are no interactions to couple the right and left moving waves (with coordinates $U$ and $V$ respectively), we can focus solely on the former. Finally, the squared Fourier coefficients}
$|\alpha_{\om \Om}|^2$ and $|\beta_{\om \Om}|^2 $ give the spectrum  
\black{$S_\mp \propto |\mA_\mp|^2$} of the positive and negative 
frequency components of Bob's purely positive frequency plane wave, as perceived by Rob.
%
%\bigskip
%Thus, we are interested in the spectrum $S_\mp(\Om)$ defined by
%\bea{S:Om}
%S_\mp(\Om) = 
%\left| 
%\int_{-\infty}^{\infty} dt\, e^{\mp i \om U}\, e^{i \Om t} 
%\right|^2
%%
%=\left| 
%\int_{-\infty}^{\infty} dt\, \ e^{i \Om t\, \mp\, i\,\left(\tfrac{\om}{a_0}\right)\,e^{-a_0 t }} 
%\right|^2,
%\eea
%where the Fourier transform has been performed with respect to Rob's proper time $t$.
The  integrals can be performed by the change of variable $y=e^{-a_0 t}$ so that
$t=-1/a_0 \ln y$, and we then write $e^{i \Om t} = y^{-i \Om/a_0}$. We can then use the Gamma function integral formula (see p50 of Keifer \cite{Kiefer:1999}, and integral (3.381.5), p318 of Gradshteyn and Ryzhik \cite{Gradshteyn_Ryzhik:1965}) 
\bea{Keifer:p50}
\hspace{-0.25in}
\int_{y=0}^{\infty} dy\, y^{\nu-1}\, e^{-(A + i\,B)\,y} = \frac{\G(\nu)}{(A+ i B)^{\nu}} =
 \frac{\G(\nu)}{(A^2 + B^2)^{\nu/2}} \, e^{-i\,\nu\,\tan^{-1}(B/A)}
\;\;\overset{A\to 0}{\to}\;\;  \G(\nu)\,|B|^{-\nu} \, e^{-i\,\nu\,\tfrac{\pi}{2}\,\trm{sign}(B)},\quad
\eea
where $\G(\nu) = \int_{0}^{\infty} dy\, y^{\nu-1}\, e^{-y}$ is the Gamma function.
One then has $\nu = -i\tfrac{\Om}{a_0}$, $A=0$ and $B = \mp \om/a_0$ for $S_\mp$, leading to
\bea{Planck:spectrum}
\black{\mA_\mp} &=& \G(\tfrac{-i\Om}{a_0}) e^{\mp\,\tfrac{\pi \Om}{2\,a_0} }\,e^{i \big(\tfrac{\Om}{a_0}\big) \ln(\om/a_0)},\no
\Rightarrow |\beta_{\om \Om}|^2 &\propto& |\G(\tfrac{-i\Om}{a_0})|^2 e^{-\tfrac{\pi \Om}{\,a_0} } 
= \tfrac{2 \pi}{(\Om/a_0)}\, \frac{1}{e^{2 \pi \Om/a_0}-1}
\equiv  \tfrac{2 \pi}{(\Om/a_0)}\, \frac{1}{e^{\hbar \Om/(k_b T_U)}-1},  \label{Planck:spectrum:line2}\\
\trm{using}\quad |\G(\nu)|^2 &=& \frac{\pi}{\nu\,\sinh(\pi\,\nu)},
\quad\trm{and defining}\quad
k_b T_U \defn \frac{\hbar\,a_0}{2\,\pi\,c}. \label{Planck:spectrum:line3}
\eea
after re-inserting the dimensional constants $k_b$ and $c$ in the Unruh temperature $T_U$, which we see by 
\Eq{Planck:spectrum:line3} has the thermal Planck spectrum for bosons. The lesson here is that 
\Eq{Planck:spectrum:line3}   reveals that Rob does \tit{not} perceive Bob's purely positive frequency plane wave as a purely positive frequency Rindler plane wave, and the latter interprets the negative frequency component spectrum $|\beta_{\om \Om}|^2 \ne 0$ as a bath of thermal particles with temperature proportional to his acceleration $a_0$.

%====================================================================
\section{Inertial-to-Rindler (I2R) coordinates}\label{sec:I2R}
%====================================================================
We now turn to the main focus of this work, the derivation and exploration of the 
Inertial-to-Rindler (I2R) coordinates. We return to general 4-velocity written as in the previous section as 
$u|_{x=0}^\mu = (T'(t), X'(t))\defn (\g(t), \g(t)\beta(t))$. 
This time, we require the velocity profile to be given by
\be{Tprime:Xprime:I2R}
\hspace{-0.75in}
\beta(t) = \b_0\,\tanh(a_0 t) 
\;\Rightarrow\;
\big(T'(t), X'(t)\big) = 
\frac{ \big(\g_0\,\cosh(a_0 t),\, \b_0\,\g_0\,\sinh(a_0 t)\big)}{\sqrt{\g_0^2 + \sinh^2(a_0 t)}},  \;\;
0\le \beta_0 \le 1,
\ee
with again $\g_0 = 1/\sqrt{1-\beta^2_0}$.
These equations are readily integrated (using a change of variable $y=\sinh(a_0 t)/\g_0$) to yield 
the trajectory of the Irwin at his origin with coordinates $(t, x=0)$
\be{T:X:I2R}
T(t) = \frac{\g_0}{a_0}\,\sinh^{-1}\left(\frac{1}{\g_0}\, \sinh(a_0 t)\right), \qquad
X(t) = \frac{\b_0\g_0}{a_0}\,\sinh^{-1}\left(\frac{1}{\b_0\g_0}\, \cosh(a_0 t)\right).
\ee
It is worth recalling the identity
\be{arcsinh:ln}
\hspace{-0.35in}
\sinh^{-1}(y) = \ln\left(y + \sqrt{y^2+1} \right), \quad\trm{with limits}\quad
\sinh^{-1}(y) \overset{y\ll 1}{\to} y, \qquad 
\sinh^{-1}(y) \overset{y\gg 1}{\to} \ln(2 y).
\ee
The equality in \Eq{arcsinh:ln} is easily proven by taking the $\sinh$ of both sides, and noting that
$1/(\sqrt{y^2+1} + y) =  \sqrt{y^2+1} -y$. 

Therefore, when $\b_0\to1$ and $\g_0\gg 1$, the argument of the 
the arcsinh is small for fixed $a_0\, t$, and we can use its small argument approximation in
\Eq{arcsinh:ln} to obtain the Rindler uniform acceleration trajectory
 $T(t)\overset{\b_0\to 1}{\to} \tfrac{1}{a_0} \sinh(a_0 t)$ and
$X(t)\overset{\b_0\to 1}{\to} \tfrac{1}{a_0} \cosh(a_0 t)$, where the factors inside and outside the arcsinh have canceled. 

In the opposite extreme, when $\b_0\to 0$ and $\g_0\to 1$ we trivially have
 $T(t)\overset{\b_0\to 1}{\to} \frac{1}{a_0}\sinh^{-1}(\sinh(a_0 t)) = t$. For the limit of $X(t)$ we note that the argument of the arcsinh is now large as $\b_0\to 0$ and so 
$X(t)\overset{\b_0\to 0}{\to} \frac{\b_0}{a_0}(\ln(2\,\cosh(a_0 t)) -\ln\b_0) \to \frac{\b_0}{a_0}\ln(e^{a_0 t}\,(1 + e^{-2a_0 t})) \approx $ $\frac{\b_0}{a_0} ( a_0 t +e^{-2a_0 t})\to \b_0\,t = \b_0\,T$ for $t\gg \tfrac{1}{a_0}$, and we have used $\b_0\,\ln\b_0\overset{\b_0\to 0}{\to} 0$, and $\ln(1 + e^{-2a_0 t}) \approx e^{-2a_0 t}$. 
Therefore in the limit $\b_0\to 0$ we obtain the inertial constant velocity trajectory of Bob, $X = \b_0\, T$.
\black{
Note that even in this limit of $\beta_0\to 0$ there is always a finite, non-zero (but in the limit, vanishing small) period of acceleration so that the I2R trajectory can turnaround from 
its asymptotically (in time) \tit{in} terminal velocity of $-\beta_0$, to its asymptotically \tit{out} terminal velocity of $\beta_0$.
} % end red

It is straightforward to compute Irwin's acceleration $a(t) = \sqrt{a^\mu(t)\,a_\mu(t)}$ by computing 
\bea{at}
a^\mu|_{x=0} &=& \big(T''(t), X''(t)\big) = a(t)\, 
\frac{\big(\b_0 \g_0 \sinh(a_0 t), \g_0 \cosh(a_0 t) \big)}{\sqrt{\g_0^2 + \sinh^2(a_0 t)}}\, \label{at:line1} \\
a(t) &=&  \frac{(\b_0\,a_0)\, \g_0^2}{\g_0^2 + \sinh^2(a_0 t)}, \qquad a(0) = \b_0\,a_0, \label{at:line2}
\eea
where the vector in \Eq{at:line1} is normalized to $(-1)$, appropriate for a spacelike 4-vector, in the metric $ds^2 = dT^2-dX^2$. 

Another interesting feature of this acceleration $a(t)$ is that upon integration we obtain
\be{chi}
\chi(t) \defn \int dt\, a(t) = \tanh^{-1}\big(\b_0 \tanh(a_0 t)\big) \quad\Rightarrow\quad 
\beta(t) = \tanh(\chi(t)),
\ee
which is a generalization of the Rindler rapidity relation $\beta_0 = \tanh(r)$, reducing to it for uniform acceleration when $\chi(t)= a_0\, t$. Using $\chi(t)$ one obtains a simple form for the 4-velocity
$u^\mu(t)=\big(T'(t), X'(t)\big) = \big(\cosh\chi(t),\sinh\chi(t) \big)$ which (i) reduces to the previous form using
the identity $\cosh(\tanh^{-1}(y)) = 1/\sqrt{1-y^2}$, and (ii) trivially yields the 4-acceleration 
$a^\mu(t)=\big(T''(t), X''(t)\big) = a(t)\, \big(\sinh\chi(t),\cosh\chi(t) \big)$. 
In fact, one could in general  have  started out using this form of the 
4-velocity for \tit{arbitrary}  $a(t)$ to produce a  4-acceleration with arbitrary proper acceleration $a(t)$. 
The difficulty is finding an expression for $a(t)$ such that 
$x^\mu (t)=\big(T(t), X(t)\big) = \big(\int dt\, \sinh\chi(t), \int dt\, \cosh\chi(t) \big)$ yields a closed form expression (although, it can be readily handled numerically).

The orbit of Irwin at his origin $x=0$ is given by solving for $\sinh(a_0\, t)$ and $\cosh(a_0 t)$
and using $\cosh^2(a_0\, t) - \sinh^2(a_0\, t)=1$ to obtain
\bea{Irwin:orbit}
&{}& \b_0^2 \g_0^2 \, \sinh^2\left(\frac{a_0 X}{\b_0 \g_0}\right)
- \g_0^2 \, \sinh^2\left(\frac{a_0 T}{ \g_0}\right)=1, \label{Irwin:orbit:line1} \\
&{}& T(t) = \frac{\g_0}{a_0}\, \sinh^{-1} \left(\beta(t)\,\sinh\left(\frac{a_0 X(t)}{\b_0 \g_0} \right) \right), \label{Irwin:orbit:line2}
\eea
where \Eq{Irwin:orbit:line1} generalizes the Rindler hyperbola $X^2-T^2 = 1/a_0^2$, and 
 \Eq{Irwin:orbit:line2} is a parametric form of $T = T(X)$ (with parameter $t$, Irwin's proper time) that will be used subsequently to define surfaces of simultaneity when we discuss radar time.
 
 Finally, to define I2R coordinates $(t,x)$ over all of Minkowski spactime, we follow the same procedure as in \Eq{Gelis:p341} for the Rindler case, namely $T - T(t) = x\,X'(t)$ and $X-X(t) = x\, T'(t)$ (i.e. the LT now involves the instantaneous velocity $\beta(t)$, vs $\b_0$ for the Rindler case) yielding

\bea{t:x:I2R}
T(t) &=& \frac{\g_0}{a_0}\,\sinh^{-1}\left(\frac{1}{\g_0}\, \sinh(a_0 t)\right) + 
x\, \frac{\b_0 \g_0 \sinh(a_0 t)}{\sqrt{\g_0^2 + \sinh^2(a_0 t)}}\, \qquad \label{t:x:I2R:line1}\\
X(t) &=& \frac{\b_0\g_0}{a_0}\,\sinh^{-1}\left(\frac{1}{\b_0\g_0}\, \cosh(a_0 t)\right) + 
x\, \frac{\g_0 \cosh(a_0 t)}{\sqrt{\g_0^2 + \sinh^2(a_0 t)}}, \label{t:x:I2R:line2}
\eea
which reduce to \Eq{Gelis:line1} and \Eq{Gelis:line2} in the limit $\b_0\to 1, \g_0\to\infty$.

Now by forming $dT = (dT/dt)\,dt + (dT/dx)\, dx$ and  $dX = (dX/dt)\,dt + (dX/dx)\, dx$ and
$ds^2 = dT^2-dX^2$ we obtain the metric analogous to the Rindler metric \Eq{Rindler:metric}
\be{I2R:metric}
ds^2_{I2R} = (1 + a(t) \,x)^2\,dt^2 - dx^2,
\ee
which effectively has the Rinlder constant proper acceleration replaced by the instantaneous I2R proper acceleration, $a_0\to a(t)$. This simple result occurs because $g_{xx} = (T')^2 - (X')^2 = u^\mu\,u_\mu = 1$ and $g_{tx} = T'\, T'' - X'\,X'' =  T'\, (a(t)\, X') - X'\,(a(t)\, T')=0$. 
Note, however, that we \tit{cannot} continue, as in the Rindler case \Eq{Rindler:metric:conformally:flat}, to form conformally flat coordinates defined by $dx \overset{?}{=} (1 + a(t)\, x)\,d\xi$ since the coefficient 
$(1 + a(t)\, x)$ is no longer a function purely of $x$.

\black{
By definition of the I2R coordinates the velocity profile of the I2R observer is 
$\beta(t) = \b_0\, \tanh(a_0\,t) \overset{t\to\infty}{\longrightarrow} \pm \b_0$, for $0\le \b_0\le 1$. Therefore, in analogy to the Rindler trajectories (with $\b_0\equiv 1$), the I2R trajectories are confined to the ``extended wedges" bounded by the slopes $dX/dT = \pm \b_0$ for all I2R time $t$, which are ``wider" than the Rindler RRW and LRW.  From \Eq{t:x:I2R:line1} and \Eq{t:x:I2R:line2} one also observes that $X(t)/T(t) \overset{t\to\infty}{\longrightarrow} \pm \b_0$ for all I2R locations $x$. Thus, 
these I2R bounded trajectory regions are not casually disconnected, although during the time periods of effectively non-zero acceleration (i.e. when $a(t)\sim \mathcal{O}(a_0)$), the I2R orbits are within the Rindler RRW and thus temporarily causally disconnected from a ``mirror" I2R observer on a similar trajectory in the LRW. We will return to our interpretation of this observation later in the paper when we heuristically argue for  an effective temperature candidate for these I2R trajectories.
} % end of red

%====================================================================
\section{ Radar time, orbits, surfaces of simultaneity, and the Twin ``Paradox" for an I2R observer}\label{sec:RadarTime:Orbits:TwinParadox}
%====================================================================
Before we plot I2R orbits, let's first discuss the concept of \tit{radar time} (see Dolby and Gull) \cite{Dolby_Gull:2001} as illustrated in \Fig{fig:radartime}.
%====================================
\subsection{Radar Time}
%====================================
Radar time (or M\"{a}rzke-Wheeler coordinates) allows an observer traveling on a worldline
$x_\G^\mu = x^\mu(\t)$ with proper time $\t$ to assign a time and distance to a remote event in spacetime, in a coordinate independent fashion, by sending a light signal to the event
and back, and averaging the sum and difference of the  proper times of the sending and
receiving times. It is clearly applicable to any observer in any gravitational background 
(see Einstein's definition in Landau and Lifshitz, \tit{Classical Theory of Fields}, p234-237)\cite{Landau_Lifshitz:CTF:1975}.
%=====================================
\begin{figure}[h!] 
\begin{center}
\includegraphics[width=2.75in,height=2.0in]{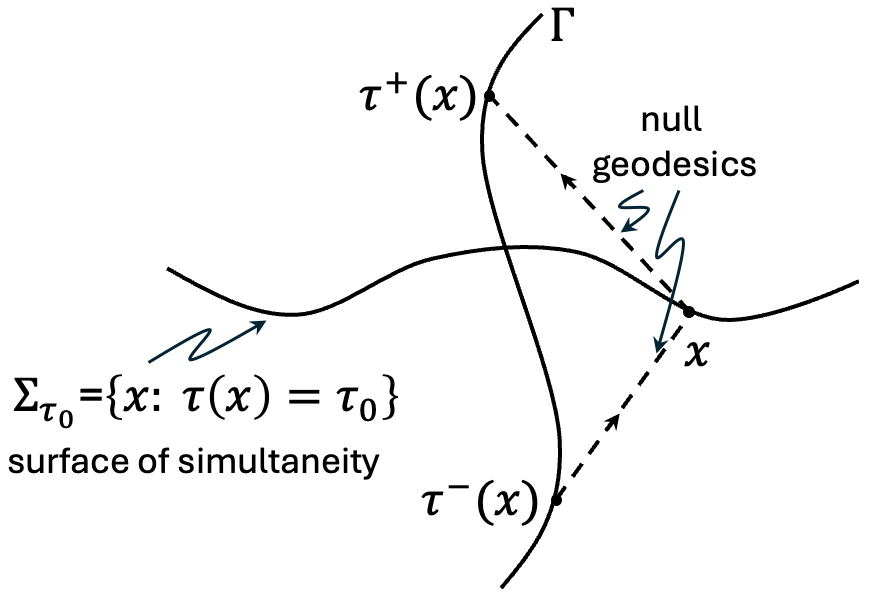} 
\end{center}
\caption{Schematic definition of \tit{radar time} $\tau(x)$.
}\label{fig:radartime}
\end{figure}
%=====================================

Consider \Fig{fig:radartime} in which an observer sends out a light signal (geodesic, traveling at $45^\circ$) at proper time $\t^{-}(x)$ that intercepts a remote spacetime event with coordinates $(\t(x), x)$, which is then immediately reflected back to the observer's worldline and received at their proper time $\t^{+}(x)$. The sending and receiving proper times are defined by (Dolby and Gull)\cite{Dolby_Gull:2001}
\begin{quote}
\begin{itemize}
\item[-] $\t^{+}(x) = $ the earliest possible proper time at which a light ray leaving the point $x$ could intercept the observer's worldline $\G$.
\item[-] $\t^{-}(x) = $ the latest possible proper time at which a light ray could leave  $\G$ and intercept the point $x$.
\item[-] $\t(x) \defn  \half (\t^{+}(x) + \t^{-}(x)) = \;$ \tit{radar time},
\item[-] $\rho(x) \defn  \half (\t^{+}(x) - \t^{-}(x)) = \;$ \tit{radar distance},
\item[-] $\Sigma_{\t_0} \defn \{x: \t(x) = \t_0 \} = \;$  observer's \tit{hypersurface of simultaneity} at their proper time $\t_0$.
\end{itemize}
\end{quote}

In their work (Dolby and Gull)\cite{Dolby_Gull:2001} provide many insightful plots, and analytic expressions, for surfaces of simultaneity for (i) Bob the constant velocity observer, (ii) Rob the Rindler uniform accelerated observer, and (iii) for a piecewise version of Irwin, who is inbound and outbound at the speed of light, but undergoes uniform acceleration for a brief period of time about the turning point of his orbit. Analytically, it is easy to illustrate the concept of radar time for the Rindler case with Rob's coordinates given by
\Eq{Rindler:coords:t:x}. We can easily write 
$T(t) = \frac{1}{a_0} \sinh(a_0\, t) =  \tanh(a_0\, t)\, X(t) \equiv \beta(t)\, X(t)$.
We therefore see that lines of Rob's constant proper time $\t_0$ are given by
$T = \beta(\t_0)\, X$ which are straight lines of constant slope $ \beta(\t_0)=\tanh(a_0\, \t_0)$ through the origin $(T,X) = (0,0)$.

%=====================================
% Original Text 28Apr2026
%=====================================
%For the I2R surfaces of simultaneity for Irwin,  we use \Eq{Irwin:orbit:line2} to write
%\be{I2R:simul}
%T = \frac{\g_0}{a_0}\, \sinh^{-1} \left(\beta(\t_0)\,\sinh\left(\frac{a_0 X}{\b_0 \g_0} \right) \right), 
%\ee
%as his surfaces of simultaneity with his fixed proper time $\t_0$, which for $\b_0\to 1, \g_0\gg1$ (and thus small arguments of the  $\sinh$,  and subsequently $\sinh^{-1}$ function) reduces to the Rindler straight lines through the origin $T = \beta(\t_0)\,X = \tanh(a_0\,\t_0)\, X$.
%In the Supplemental Material \cite{SM:I2R:30Apr2026} we show the codes for computing the radar time numerically following 
%the analytical procedure of Dolby and Gull\cite{Dolby_Gull:2001}, which produced the contour plots shown in \Fig{fig:oribits:lines:of:simultaneity:1} and \Fig{fig:oribits:lines:of:simultaneity:2}  below.

%=====================================
% New Text 30May2026
%=====================================
\black{
For the I2R surfaces of simultaneity for Irwin's trajectory ($x=0$), we follow the analytic procedure of Dolby and Gull\cite{Dolby_Gull:2001}, to develop a simple numerical procedure (see the \tit{Mathematica} codes in the Supplemental Material \cite{SM:I2R:30Apr2026}), thus avoiding the enumeration of numerous analytical special cases and regions,
 used to produce the contour plots shown in \Fig{fig:oribits:lines:of:simultaneity:1} and \Fig{fig:oribits:lines:of:simultaneity:2}  below.
We first grid up a region of Minkowski spacetime with coordinates $(T_0,X_0)$, and form the lightcone coordinates $U_0\defn X_0+T_0$ and $V_0\defn X_0-T_0$. 
Then for the comoving I2R observer with fixed $(a_0, \beta_0)$, 
we solve the following equations for $t$, using
\Eq{T:X:I2R}, for each pair of points $(T_0, X_0)$ to yield an associated radar time 
$\t = \t(T_0,X_0)$ for the point $(T_0, X_0)$.
\bea{SufSim:Eqns}
\trm{Solve for t }\ni\, \quad X(t; a_0, \b_0) + T(t; a_0, \b_0) &=& U_0 \defn X_0 + T_0, \quad\Rightarrow\quad t\to t^+, \no
\trm{Solve for t }\ni\, \quad X(t; a_0, \b_0) - T(t; a_0, \b_0) &=& V_0 \defn X_0 - T_0, \quad\Rightarrow\quad t\to t^-, \no
\Rightarrow \t^+ = \trm{max}[t^+, t^-], \quad \t^- &=& \trm{min}[t^+, t^-], \quad \t = \half\, (t^+ + t^-).
\eea
We then simply use the \ttt{ContourPlot} feature in \tit{Mathematica} to automatically connect  lines of simultaneity with the same value $\t$ throughout the initial gridded Minkowski spacetime region.
} % end of red

%=====================================
\begin{figure}[h!]
\begin{center}
\begin{tabular}{ccc}
\includegraphics[width=2.25in,height=1.75in]{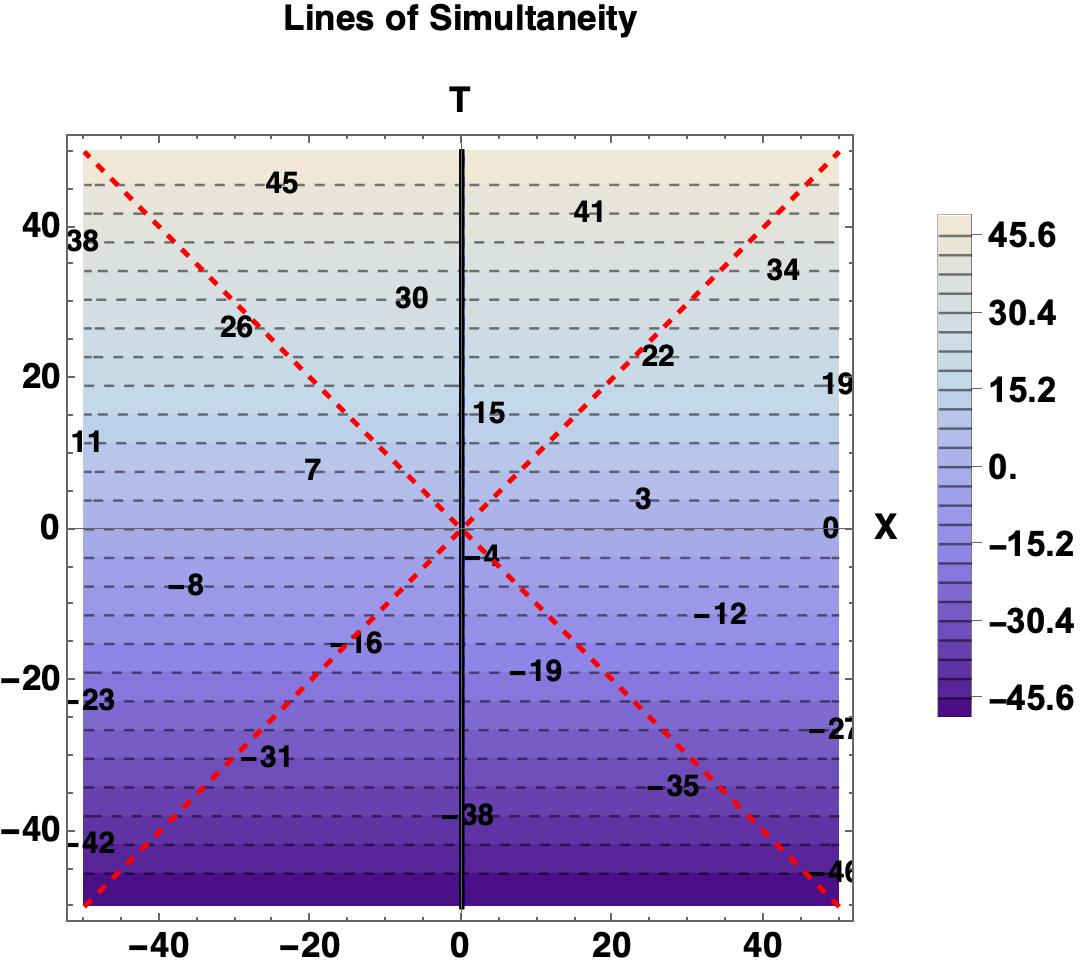} &
\includegraphics[width=2.25in,height=1.75in]{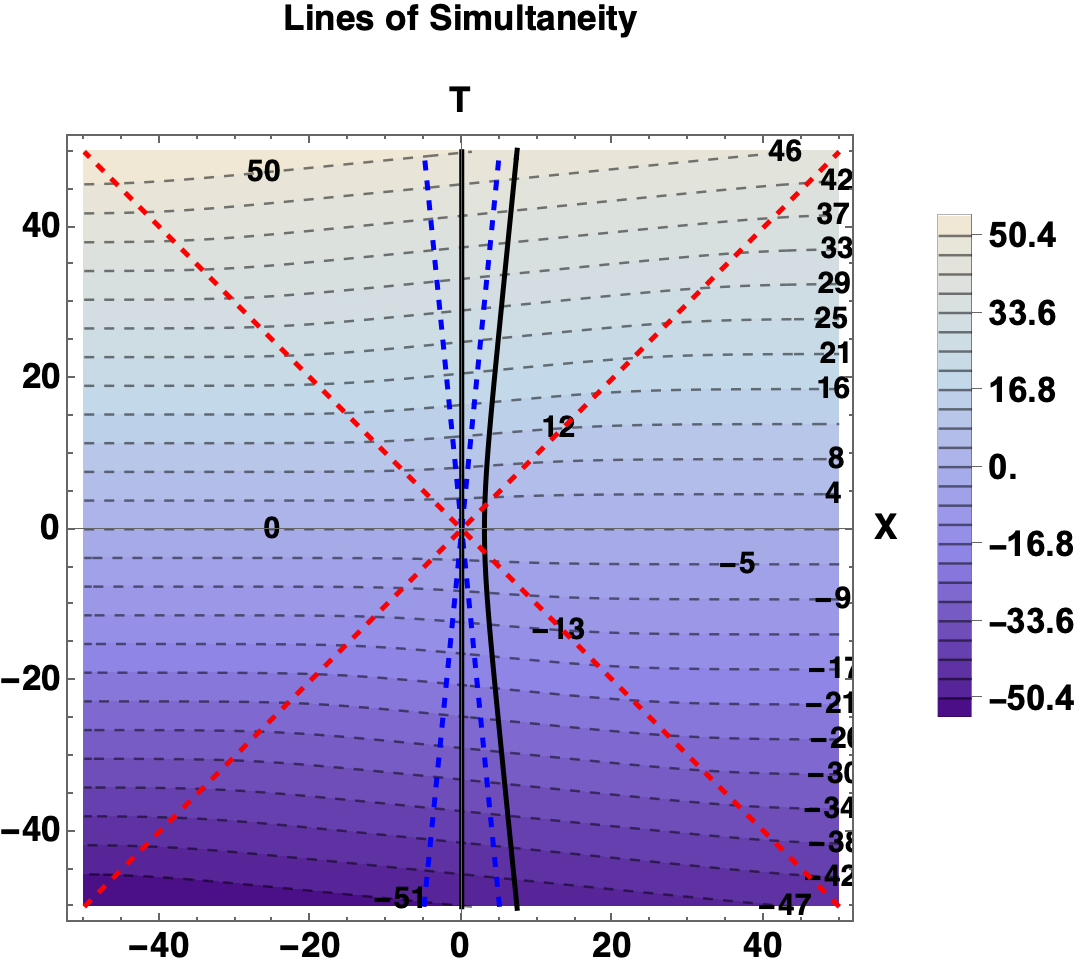} & 
\includegraphics[width=2.25in,height=1.75in]{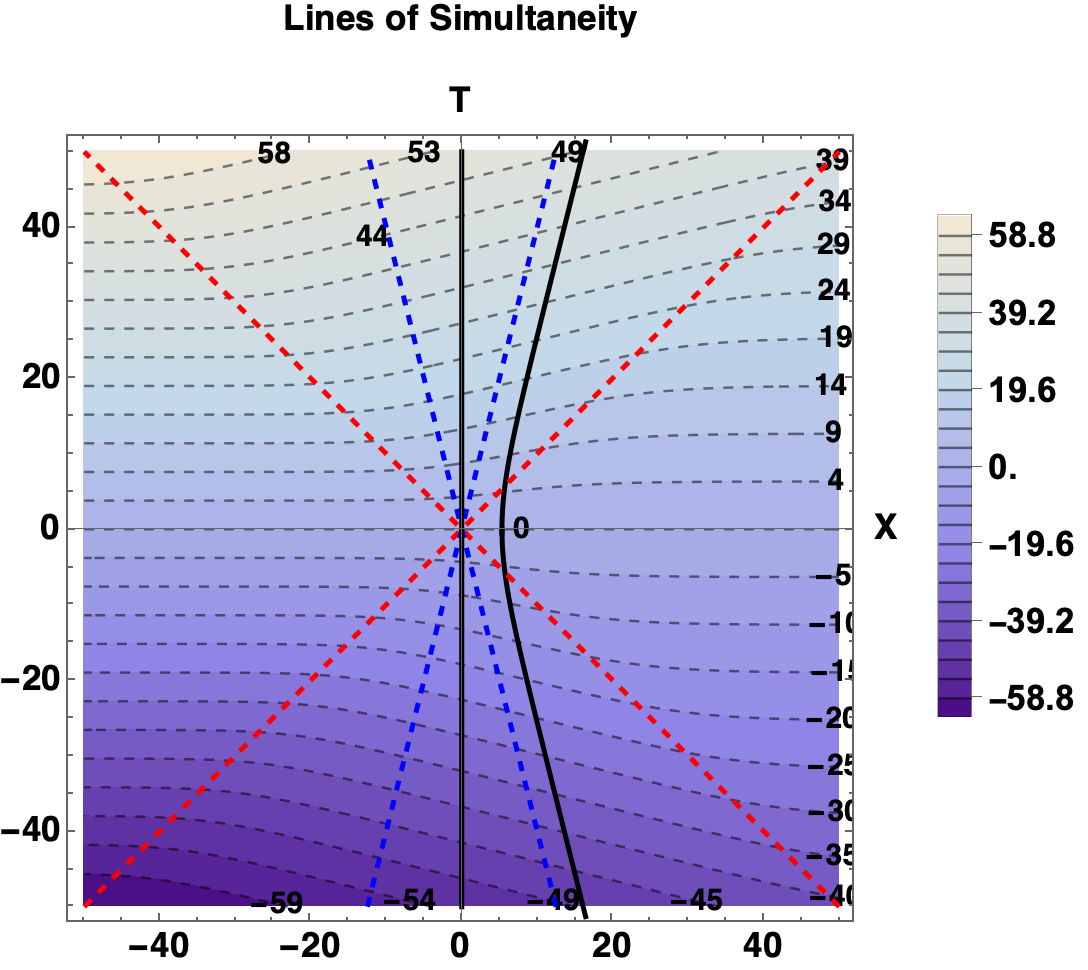}
\end{tabular}
\end{center}
\caption{Orbits and lines of simultaneity $\t_0$  for $a_0 = 0.1$ and $\b_0 = \{0.0001, 0.1, 0.25\}$.
(solid~black)~$I2R$ orbit (Irwin), (dashed red) lightcone $T=\pm X$, 
(dashed blue) Irwin's asymptotic velocity $T=\pm \b_0 X$, 
(dashed gray)  lines of simultaneity $\t_0$ (labeled).
}\label{fig:oribits:lines:of:simultaneity:1}
\end{figure}
%=====================================
%=====================================
\begin{figure}[h!]
\begin{center}
\begin{tabular}{ccc}
\includegraphics[width=2.25in,height=1.75in]{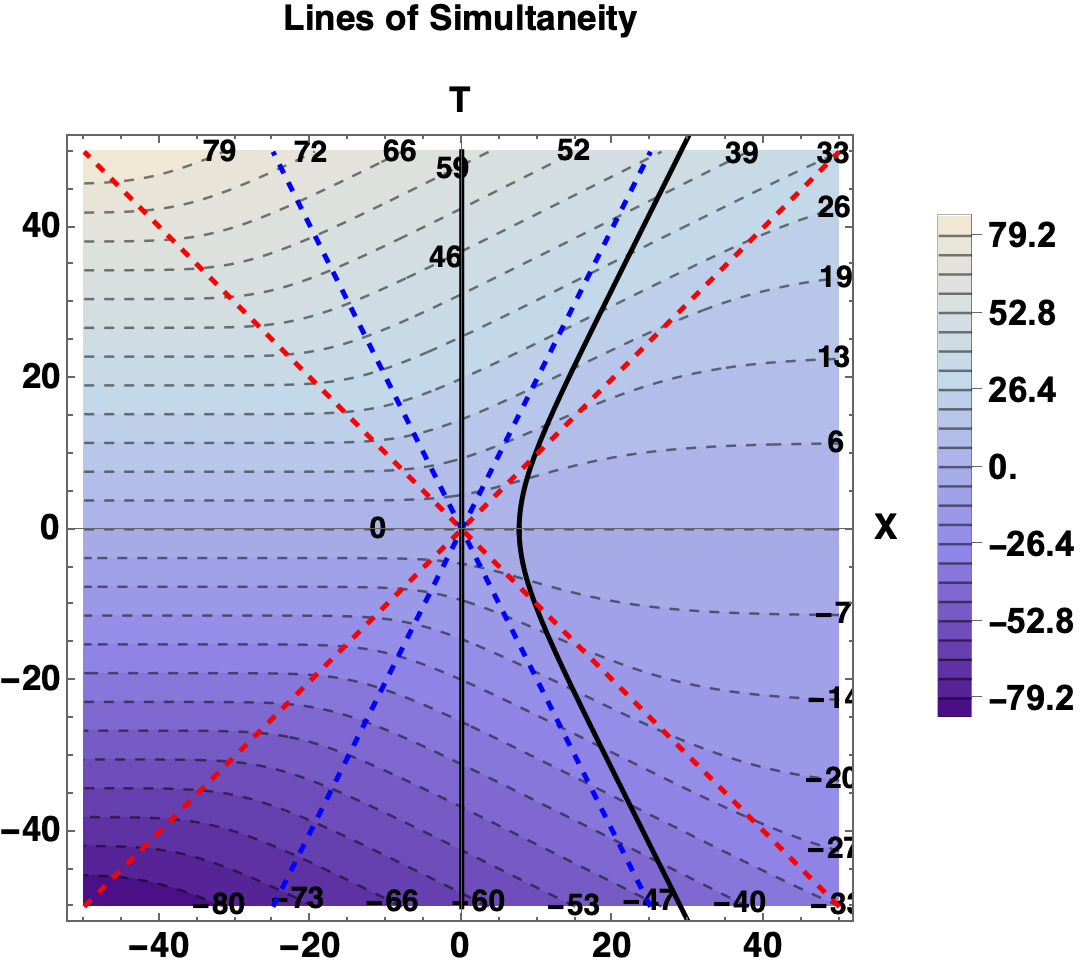} &
\includegraphics[width=2.25in,height=1.75in]{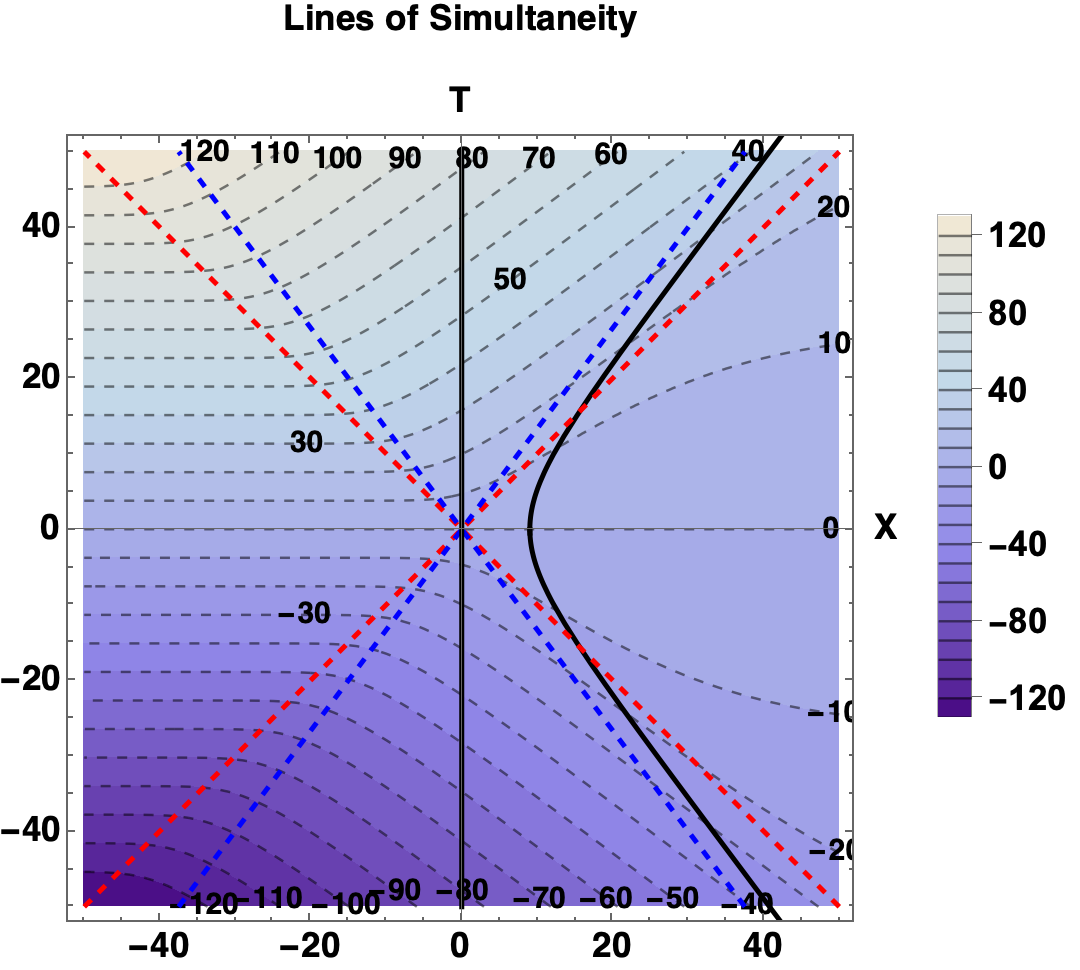} & 
\includegraphics[width=2.25in,height=1.75in]{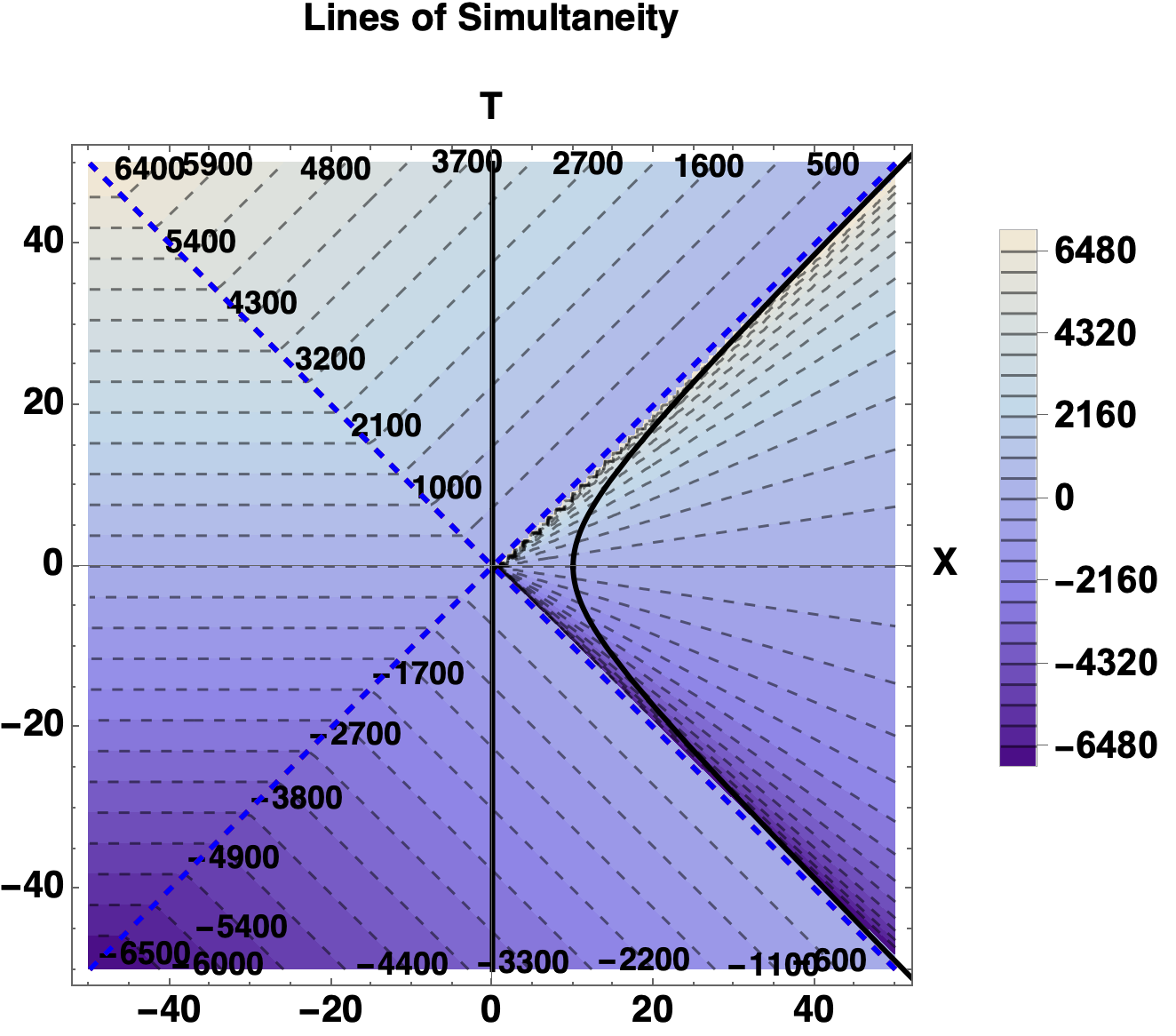}
\end{tabular}
\end{center}
\caption{Orbits and lines of simultaneity $\t_0$ for $a_0 = 0.1$ and $\b_0 = \{0.5, 0.75, 0.9999\}$.
(solid~black)~$I2R$ orbit (Irwin), (dashed red) lightcone $T=\pm X$, 
(dashed blue) Irwin's asymptotic velocity $T=\pm \b_0 X$, 
(dashed gray)  lines of simultaneity $\t_0$ (labeled).
}\label{fig:oribits:lines:of:simultaneity:2}
\end{figure}
%=====================================

%====================================
\subsection{Orbits and Surfaces of Simultaneity}
%====================================
In the figures below we see that the limits we discuss previously are borne out, namely as $\b_0\to 0$
\Fig{fig:oribits:lines:of:simultaneity:1} (left) we have essentially inertial orbits $T = \pm \b_0 X$, while
for $\b_0\to 1$ \Fig{fig:oribits:lines:of:simultaneity:2} (right) we have a Rindler uniform acceleration orbit 
completely confined to the RRW. For intermediate values of $0<\b_0<1$ Irwin's orbit remains inside the RRW for a time that depends on both $(a_0, \b_0)$.
\Fig{fig:oribits:lines:of:simultaneity:1}(middle, right) for low values of the velocity $\b_0=\{0.1,0.25\}$, are reminiscent  of the ``immediate turn-around plot Fig. (5) of Dolby and Gull \cite{Dolby_Gull:2001},
where close to Irwin's trajectory the lines of simultaneity are tilted positively (negatively)  for the outbound (inbound) portion of the trajectory with $t>0 (<0)$, with slope $\b_0 (-\b_0)$ with respect to Bob's $X$-axis (i.e. parallel to Irwin's $x$-axis), and flatten out to zero slope as they enter the RRW. 
For larger values of 
$\b_0 = \{0.5, 0.75\}$ \Fig{fig:oribits:lines:of:simultaneity:2}(left, middle) are similar to their "gradual turn-around" plot, Fig.(6)\cite{Dolby_Gull:2001} where there is significantly more curvature in the lines of simultaneity as they pass through Irwin's trajectory and enter the RRW, and flatten out to zero slope there. 
Finally, near the speed of light value $\b_0 = 0.9999$, the RRW of \Fig{fig:oribits:lines:of:simultaneity:2}(right) is 
reminiscent of the  uniformly accelerating Rindler case in  Dolby and Gull's
Fig.(7)\cite{Dolby_Gull:2001} (they only show the RRW). In the latter, we also show the lines of simultaneity \tit{outside} the RRW, in which the horizontal lines of constant $\t_0$ in the LRW bend sharply in the future and past regions to be parallel to the asymptotes $T=\pm X$, and are hence ``repelled" by the RRW, the latter with their own lines of simultaneity through the origin, $T = \b_0\tanh(a_0\, \t^*)\, X$ with $\b_0\approx 1$.

\begin{figure}[h!]
\begin{center}
\begin{tabular}{ccc}
\includegraphics[width=2.25in,height=1.75in]{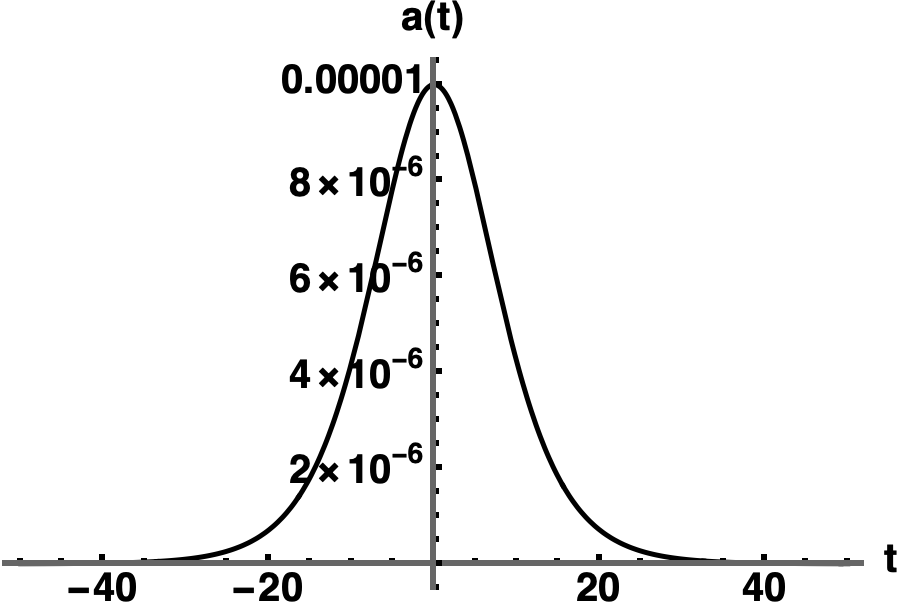} &
\includegraphics[width=2.25in,height=1.75in]{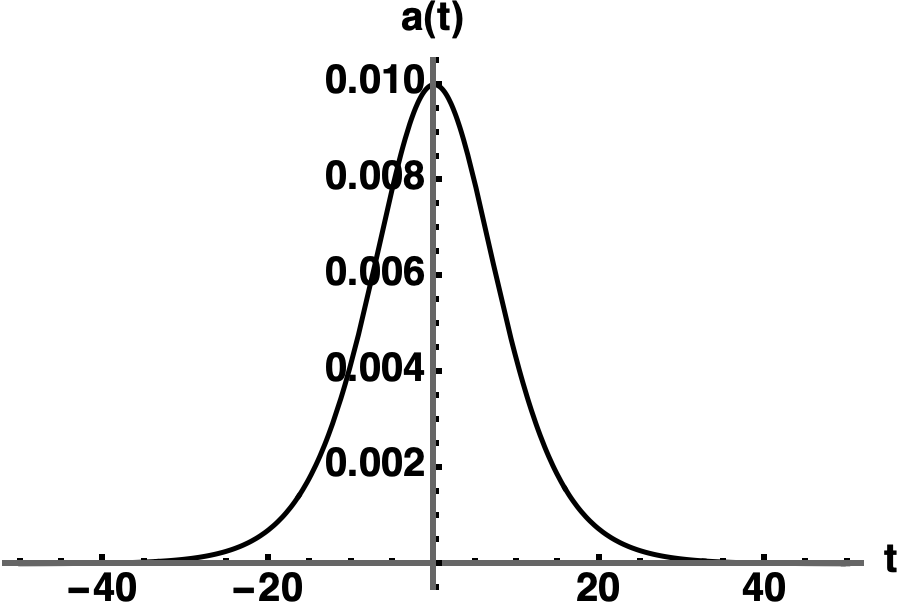} & 
\includegraphics[width=2.25in,height=1.75in]{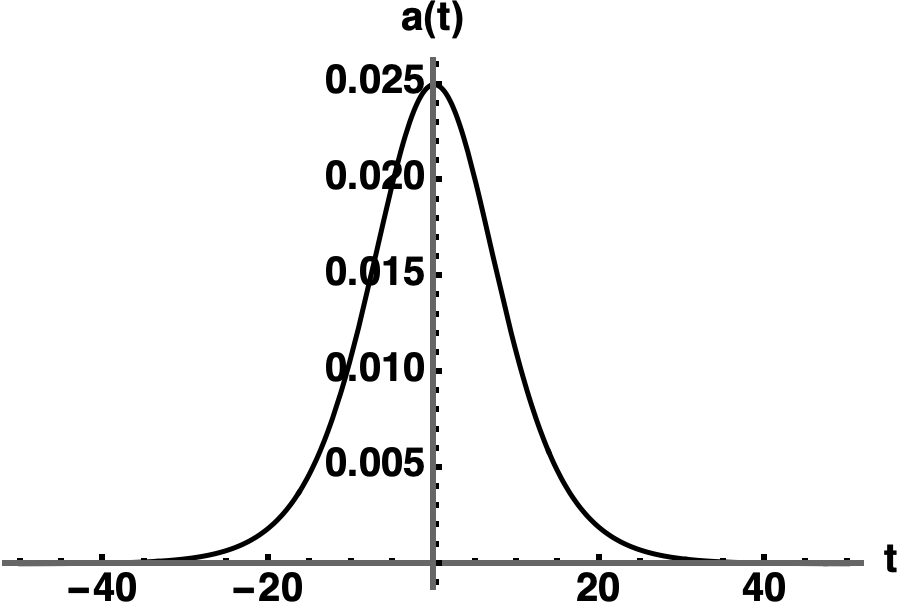}
\end{tabular}
\end{center}
\caption{Acceleration profiles for $a_0 = 0.1$ and $\b_0 = \{0.0001, 0.1, 0.25\}$.
}\label{fig:acceleration:profile:1}
\end{figure}

\begin{figure}[h!]
\begin{center}
\begin{tabular}{ccc}
\includegraphics[width=2.25in,height=1.75in]{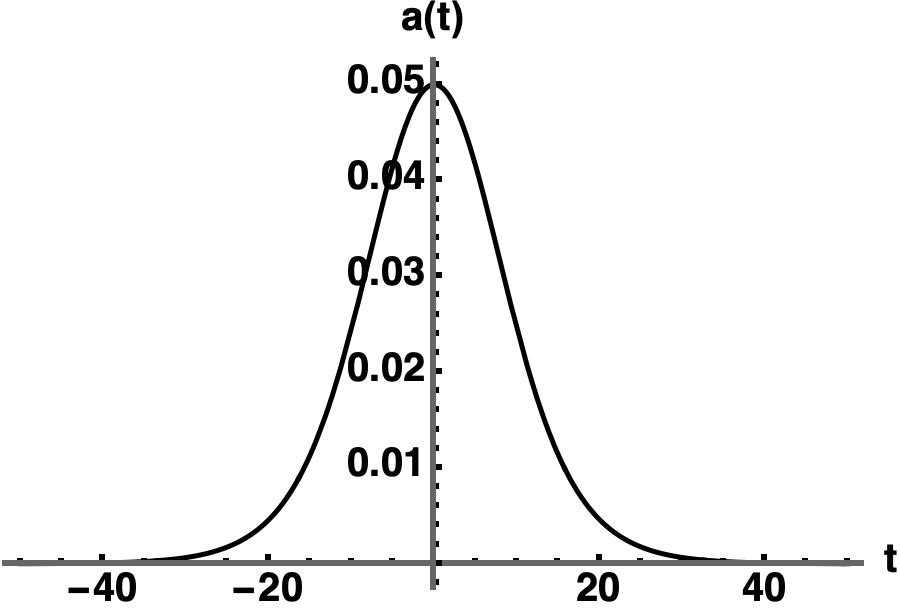} &
\includegraphics[width=2.25in,height=1.75in]{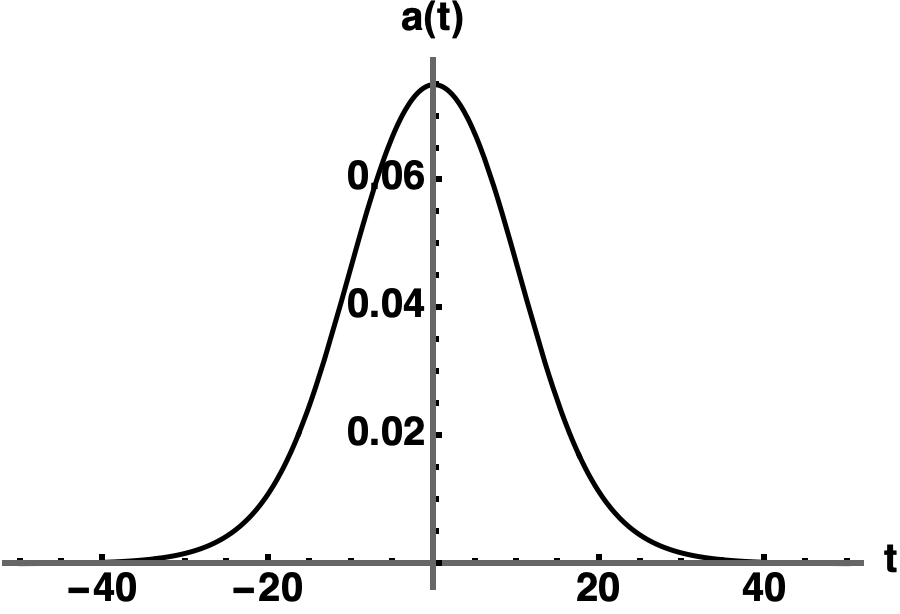} & 
\includegraphics[width=2.25in,height=1.75in]{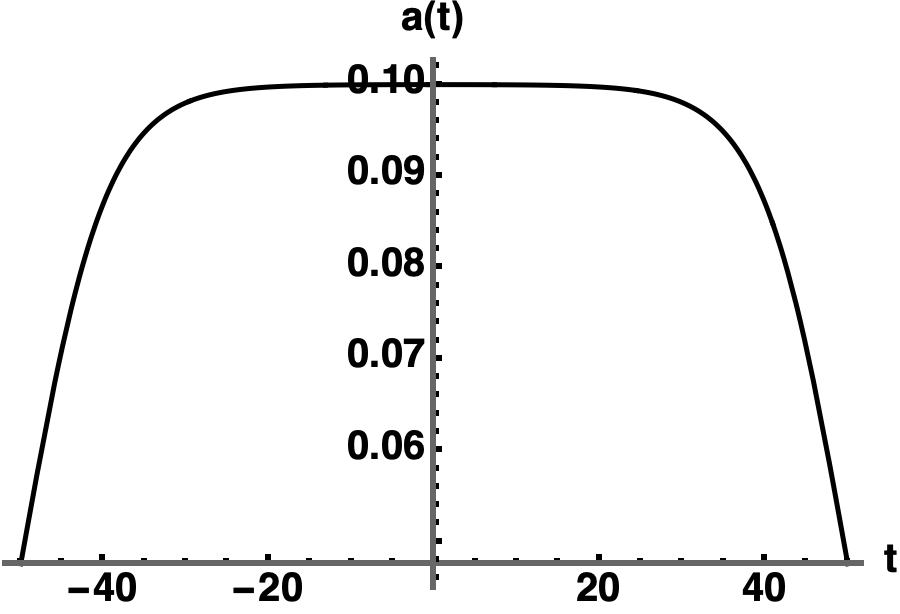}
\end{tabular}
\end{center}
\caption{Acceleration profiles for $a_0 = 0.1$ and $\b_0 = \{0.5, 0.75, 0.9999\}$.
}\label{fig:acceleration:profile:2}
\end{figure}

The acceleration profiles are shown in \Fig{fig:acceleration:profile:1} and \Fig{fig:acceleration:profile:2}.
%=====================================
%=====================================
%=====================================
%=====================================
For low values of $\b_0$, they have similar shapes, but different scaling, with slowly widening widths.
It is only when the begining/end terminal velocity is extremely relativisitic $\b_0\gtrsim 0.95$ for this chosen value of $a_0=0.1$, do we see the acceleration profile really begin to widen to $\approx \b_0\,a_0$ over a broad range of proper time.

\begin{figure}[h!]
\begin{center}
\includegraphics[width=2.75in,height=2.0in]{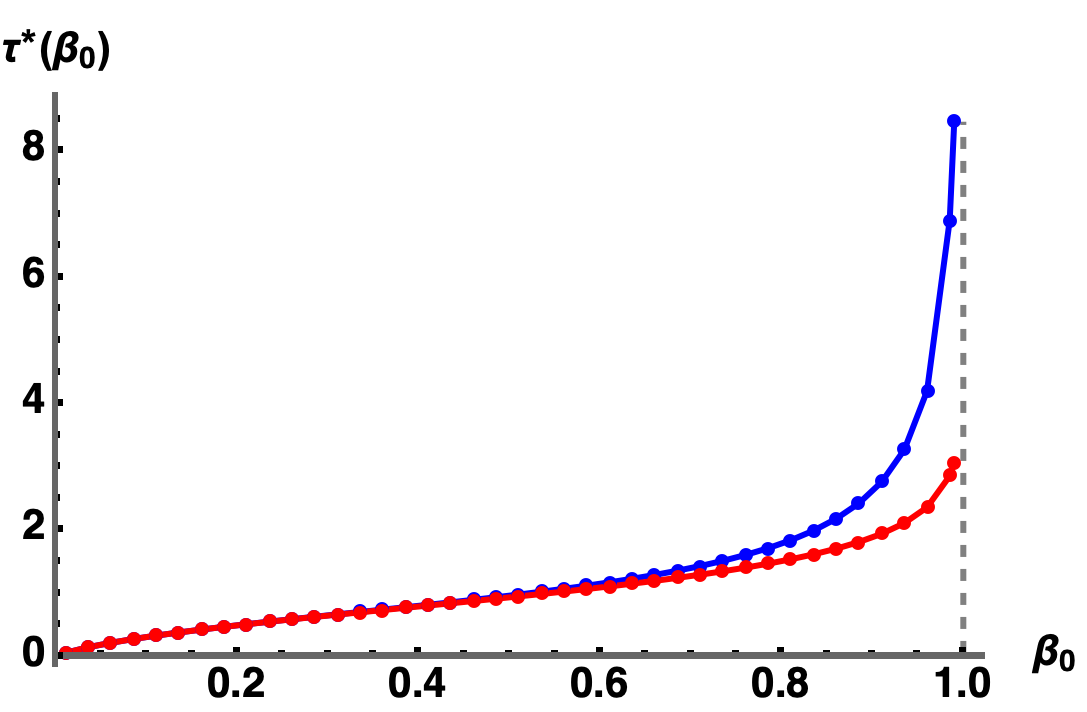}
\end{center}
%\caption{Scaled proper time (blue) Bob's, (red) Irwin's,
%that Irwin's orbit is completely inside \black{Rob's} RRW, $X>|T|$.
%}
\caption{\black{(Color online)  
(blue) Scaled proper time $\tau^*(\beta_0)$ that Bob perceives  Irwin 
to be inside Rob's RRW , $X>|T|$, with positive velocity.
(red) Irwin's scaled proper time to be  inside Rob's RRW with positive velocity.
}% end red
}
\label{fig:T:t:Irwin:in:RRW}
\end{figure}

The time Irwin spends in the RRW with positive velocity is 
given by the (scaled) Minkowski  time $\t^*_T = a_0 T$, defined when his orbit intercepts the lightcone $T=X$.
This time is obtained by solving Irwin's orbit equation \Eq {Irwin:orbit:line1} with $X\to T$.
This is plotted as the blue curve in \Fig{fig:T:t:Irwin:in:RRW} below as a function of $\b_0$.
One then finds Irwin's associated proper time $\t^*_t$ by inverting the I2R coordinate $T(t)$ for $t$ in \Eq{T:X:I2R}. This is plotted as the red curve in  \Fig{fig:T:t:Irwin:in:RRW}. 
%=====================================
%=====================================
We see that for $\b_0\lesssim \tfrac{1}{\sqrt{2}}\sim 0.707$ (at which point $\sinh(a_0\, \t^*) = \g_0$, and $a(a_0\,\t^*)=\half a(0)$) the two curves are approximately the same $\t^*_T\approx \t^*_t$, and roughly linear in $\b_0$ over $0\le t \le \t^*$. However, after this time $\t^*$, both curves begin to rise exponentially, with Irwin's orbit approaching more closely to a Rindler-like constant acceleration orbit as $\b_0\to 1$.

%=====================================
\subsection{Revisiting the Twin ``Paradox"}
%=====================================
Let us now revisit the classic Twin ``Paradox." 
Let Bob be an Earthbound inertial observer with coordinates $(T,X)$ recording the passage of time for a 
traveling twin, with coordinates $(t,x=0)$, with velocity profile $\beta(t)$ relative to Bob.
Their metrics are related by $dt^2 = dT^2 - dX^2 = dT^2 \big(1- \beta^2(t)\big)$. Therefore, the time that passes for Bob during the twin's journey is given by  $T = \int_{t_i}^{t_f} \frac{dt}{\sqrt{1-\beta^2(t)}}$ (where $t$ is the twin's proper time), 
which defines the appropriate time transformation between 
Bob and  (i) Charlie (C) a constant velocity traveler, (ii) Rob (R)  a uniformly accelerating traveler, and 
(iii) Irwin (I) an I2R traveler, with  velocity profiles $\beta(t) = \big(\b_0, \tanh(a_0 t), \,\b_0\,\tanh(a_0 t) \big)$, respectively. Further, let us define the scaled proper time of the traveler as $\t \defn a_0 t$ ($a_0\,T\to T_\t$), and consider the specific case when each traveler ages 3 years of their proper time $\t_{C,R,I}$ during half the trip, i.e. 
$\t^* = a_0\,t^* = 3$. We wish to consider the time $T_{\t_{C,R,I}}(\b_0)$ that Bob ages during one half of the twin's journey as a function of the parameter $\b_0$.
%=====================================
\begin{figure}[h!]
\begin{center}
\begin{tabular}{ccc}
\hspace{0.5in}
\includegraphics[width=1.35in,height=1.875in]{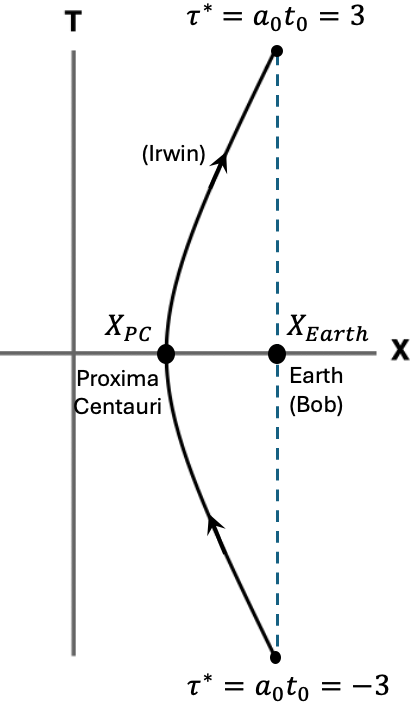} &
\hspace{0.25in} {}&
\hspace{0.5in}
\includegraphics[width=2.5in,height=1.75in]{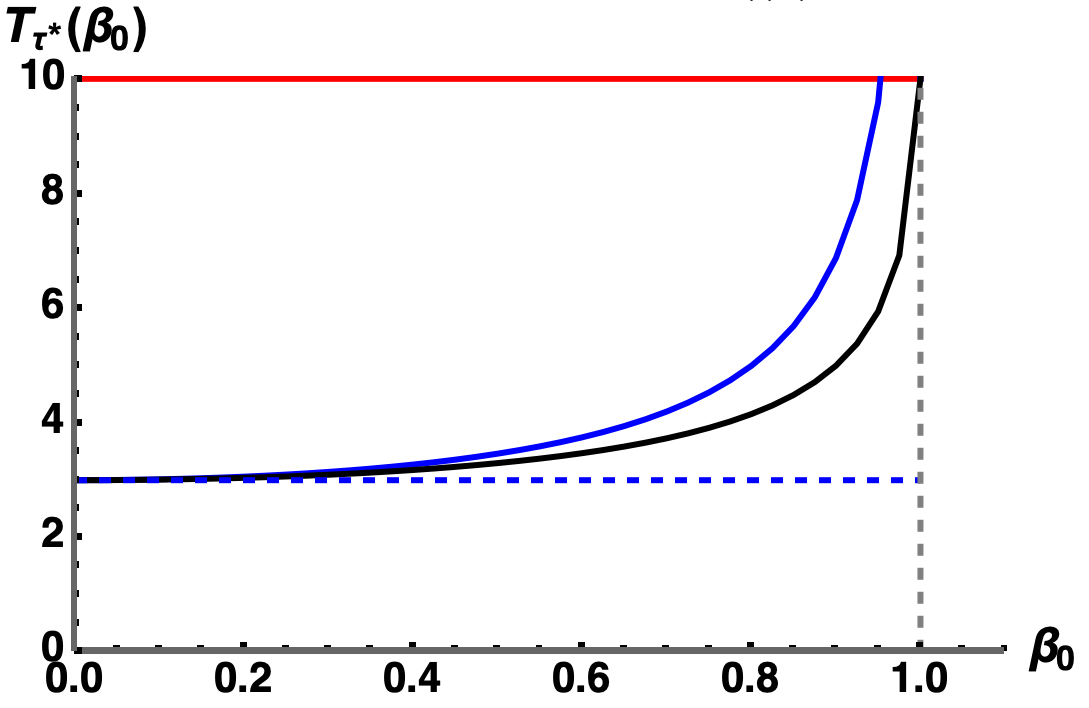}  
\end{tabular}
\end{center}
%\caption{
%(left) Twin ``Paradox" orbit: Bob (inertial, blue dashed vertical curve), Irwin (I2R, black curve).
%(right) Bob's age $T(\t^*)$ at half the traveling twin's trip for 
%(i) (blue-dashed) Bob (B) himself,
%(ii) (blue solid) Charlie (C) a constant velocity $(\b_0, a_0=0)$ traveler, 
%(iii) (red) Rob (R)  a uniformly accelerating traveler $(\b_0=1, a_0)$, and 
%(iv) (black) Irwin (I)  an I2R traveler,  with   
%velocity profiles $\beta(t) = \big(0,\, \b_0,\, \tanh(a_0\, t), \,\b_0\,\tanh(a_0\, t) \big)$,
%when each traveler records their own proper time (age) $\t^{B,C,R,I} = a_0\,t^* = 3$.
%(Note: $\sinh(3) = 10.018$).
%}
\caption{
(left) Twin Paradox orbit: Bob (inertial, blue dashed vertical curve), Irwin (I2R, black curve).
(right) \black{(Color online)}  Bob's age \black{$T_{\t^*}(\beta_0)$} at half the traveling twin's trip for 
(i) (blue-dashed) Bob (B) himself,
(ii) (blue solid) Charlie (C) a constant velocity $(\b_0, a_0=0)$ traveler, 
(iii) (red) Rob (R)  a uniformly accelerating traveler $(\b_0=1, a_0)$, and 
(iv) (black) Irwin (I)  an I2R traveler,  with   
velocity profiles
$\beta(t) = \big(0,\, \b_0,\, \tanh(a_0\, t), \,\b_0\,\tanh(a_0\, t) \big)$  \black{for (i)-(iv)},
when each traveler records their own proper time (age) $\t^{B,C,R,I} = a_0\,t^* = 3$.
(Note: $\sinh(3) = 10.018$).
}
\label{fig:twin:paradox:proper:times}
\end{figure}
%================================

Consider first Bob himself with $\b_0=0$ (or someone else stationary with respect to Bob), so that $T_{\t_{B}} = \t^* = 3$.
This is depicted as the bottom constant dashed blue line in \Fig{fig:twin:paradox:proper:times}(right).

Second, consider the inertial twin Charlie who moves with constant relative velocity $\b_0$ with respect to Bob. Then $T_{\t_{C}}  = \g_0\,\t^* = \t^*/\sqrt{1-\b_0^2}$, which is depicted as the solid blue line in \Fig{fig:twin:paradox:proper:times}(right).

For the uniformly accelerated ($a_0$) twin Rob, we have Bob's age given by the Rindler time transformation
$T_{\t_{R}} = \sinh(\t^*)$ (independent of $\b_0$, since $\b_0\equiv 1$), which is the top constant red curve in \Fig{fig:twin:paradox:proper:times}(right).

Finally, for the I2R twin Irwin \Fig{fig:twin:paradox:proper:times}(left), we have Bob's age given by the I2R time transformation $T_{\t_{I}} = \g_0\,\sinh^{-1}\big(\sinh(\t^*)/\g_0\big)$, 
which is depicted as the solid black curve 
 in \Fig{fig:twin:paradox:proper:times}(right).
 As anticipated, Bob's age  at half of Irwin's trip (black curve), as a function of $\b_0$, 
 mimics the behavior of his age with respect to that of Charlie's trip (solid blue curve) for low velocities, $\b_0$ small, while rising rapidly to merge with his age with respect to Rob's trip (red curve) as $\b_0\to 1$.
 This is just another indication of how the I2R coordinates smoothly transforms between those of the inertial, constant velocity traveler Charlie, and uniformly accelerated traveler Rob, as we vary $\b_0\in [0,1]$ (for fixed Rindler acceleration  value $a_0$). The I2R oribt smoothly incorporates both the "gradual turnaround" scenario for small, moderate 
 $\b_0$ (for some fixed $a_0$), as well as  has the limits of the ``immediate turn-around"  scenario for $\b_0\ll 1$, as well as  the constant acceleration scenario for $\b_0\to 1$. 
 
%===================================================
\section{Corrections to the Unruh temperature}\label{sec:I2R:Temp}
%===================================================
In this section we return to the calculations \Eq{alpha:beta} - \Eq{Planck:spectrum:line3}  performed in \Sec{subsec:Unruh:effect} for  the frequency content of inertial Bob's positive frequency plane wave as seen by the uniformly accelerated Rindler observer, but now adapted to an I2R observer, Irwin. We will consider two extreme limits (i) $0<\b_0 = \eps \ll 1$, and (ii) $\b_0 = 1-\eps$. 
From these results we will argue for a proposed \tit{I2R temperature} $T_{I2R}$ defined by
\be{TI2R}
\black{k_b}\,T_{I2R} \defn \frac{\hbar\,a(t)}{2 \pi c}, \qquad a(t) = \frac{(\b_0\,a_0)\,\g_0^2}{\g_0^2 + \sinh^2{a_0\,t}},
\ee
\black{where $k_b$ is Boltzmann's constant, which we will set to unity from here on out,}
which generalizes the Unruh temperature $\black{k_b}\,T_U = \tfrac{\hbar a_0}{2 \pi c}$, and which reduces to the later in the limit $\b_0\to 1$, and to zero for $\b_0\to 0$.
\black{Note that we do not propose that the effective temperature in  \Eq{TI2R} is valid for \tit{any} kind of accelerated motion; rather \tit{only} for orbits obtained from our current I2R coordinates, since they smoothly interpolate between the linear inertial trajectory $(a_0=0, \beta_0\to 0)$, and 
constant Rindler acceleration hyperbolic trajectory $(a_0, \beta_0\to 1)$. For example is know that if one has uniform circular acceleration
%(see “Electrons as accelerated thermometers” by J.S. Bell and J.M. Leinaas, Nucl. Phys. B 212, 131 (1983)), 
one can not strictly define a temperature in the same way one can for linear, uniform acceleration\cite{Letaw_Pfautsch:1980,Letaw:1981,Bell_Leinaas:1983}.}

%===================================================
\subsection{Limit $\boldsymbol{\b_0\ll 1}$}\label{subsec:I2R:Temp}
%===================================================
Let us return to the integral in \Eq{alpha:beta} and write this as
\be{A:mp}
\mA_{s=\mp 1} \propto
\begin{bmatrix}
\alpha_{\om \Om} \\
\beta^*_{\om \Om} \\
\end{bmatrix}
%
%= \frac{1}{2 \pi}\,\sqrt{\frac{\Om}{\om}}\,
\defn \frac{1}{2\pi}\int_{-\infty}^{\infty} dt\, e^{\mp i \om U}\, e^{i \Om t},
\ee
were, again, $\alpha_{\om \Om}$ and $\beta^*_{\om \Om}$ are the Fourier amplitudes for the positive/negative frequency content as perceived by Irwin, for the  purely positive frequency (right moving)  plane wave  $e^{-i \om U}$ (where $U = T-X)$ of the inertial Bob. 
For the limit we are considering $\b_0\ll 1$ we can expand the I2R time and space transformations \Eq{T:X:I2R}  to first order in $\b_0$ to obtain
\be{}
\hspace{-0.25in}
T \sim t + \mO(\b_0^2), \; 
X\sim \frac{\b_0}{a_0}\,\ln[2\,\cosh(a_0\,t)]+ \mO(\b_0^2) \;\;\Rightarrow\;\;
U\sim t- \frac{\b_0}{a_0}\,\ln[2\,\cosh(a_0\,t)]+ \mO(\b_0^2).
\ee
\black{
Let us define the scaled time $\eta = a_0\,t$, 
% and note that the second term in
%$U\sim\tfrac{1}{a_0}(\eta - \b_0\,\ln[2\,\cosh(\eta)])$ does not change sign under $\eta\to -\eta$, while the first term clearly does. 
and set $s = \mp 1$ associated with the plane wave $e^{s\, i \,\om\, U}$ in 
\Eq{A:mp}. We can then write
\bea{A:s}
\mA_s &\defn&
%= \frac{1}{2 \pi}\,\sqrt{\frac{\Om}{\om}}\,
\frac{1}{2 \pi a_0}
\int_{-\infty}^{\infty} d\eta\,e^{i \left(\tfrac{\Om}{a_0}\right)\,\eta} \, 
e^{ i\,s\,\left(\tfrac{\om}{a_0}\right)\,\big(\eta - \b_0\,\ln[2\,\cosh(\eta)]\big)}, \no
&=& 
\frac{1}{2 \pi a_0}
\int_{-\infty}^{\infty} d\eta\, e^{\tfrac{i}{a_0}\left(\Om + s\, \om\,(1- \b_0)\,
\right)\,\eta}\,
e^{\frac{-i s \om \beta_0}{a_0} \, \ln(1+e^{-2 \eta})}, \no
&\approx&
\frac{1}{2 \pi a_0}
\int_{-\infty}^{\infty} d\eta\, e^{\tfrac{i}{a_0}\left(\Om + s\, \om\,(1- \b_0)\,
\right)\,\eta}\,
\left[
1 - \frac{i s \om \beta_0}{a_0}\,\ln(1+e^{-2 \eta}) + \mathcal{O}(\beta_0^2)
\right],
\eea
where we have used 
 $\ln[2\,\cosh(\eta)] = \ln[e^\eta\,(1+ e^{-2\,\eta})] = \eta + \ln[(1+ e^{-2\,\eta})]$, and have expanded the exponential in the second line above to first order in $\beta_0$.
Let us now separately examine the two contributions of  $\mathcal{O}(1)$
and $\mathcal{O}(\beta_0)$ in the large square brackets in \Eq{A:s}.
} % end red
%
%Let us first  break up the integral into positive and negative values of $\eta$ and 
%note that $\int_{-\infty}^0 d\eta\, f(\eta) \overset{\eta\,\to\, -\eta}{\longrightarrow} \int_0^\infty d\eta\,f(-\eta)$, and again recall that under this transformation $\ln[2\,\cosh(\eta)]$ does not change sign.
%We now approx  $\ln[2\,\cosh(\eta)] = \ln[e^\eta\,(1+ e^{-2\,\eta})] = \eta + \ln[(1+ e^{-2\,\eta})] \approx \eta +  e^{-2\,\eta}$ in two ways. 

%===================================================================
\subsubsection{\black{The $\mathcal{O}(1)$ contribution to \Eq{A:s}}}
%===================================================================
\black{
For the first $\mathcal{O}((\beta_0)^0=1)$ term in the large square bracket in \Eq{A:s}
let us break up the integral into positive and negative values of $\eta$ and 
note that $\int_{-\infty}^0 d\eta\, f(\eta) \overset{\eta\,\to\, -\eta}{\longrightarrow} \int_0^\infty d\eta\,f(-\eta)$. 
%
%First, we drop the second ``chirp" term, assuming $e^{-2\,\eta}\ll \eta$, which is approximately true for most of the integration domain, and arrive at 
We then have
} % end red
\bea{A:s:first:approx}
\mA_s^{(0)} &=& \mA^{(0) +}_s + \mA^{(0) -}_s,\quad\trm{with}\quad
 \mA^{(0) \pm}_s \defn
\frac{1}{2 \pi a_0}
\int_{0}^{\infty} d\eta\,
e^{\pm\,\tfrac{i}{a_0}\left(
\Om + s\, \om\,(1\mp \b_0)
\right)\,\eta}, \no
\Rightarrow\; 
\begin{bmatrix}
\alpha_{\om \Om} \\
\beta^*_{\om \Om} \\
\end{bmatrix}
&\propto&
\mA_{s=\mp 1}^{(0)} 
=\frac{1}{2}
\left[
\d\big(\Om \mp \om (1- \b_0)\big) + \d\big(\Om \mp \om (1 + \b_0)\big), 
\right]
\eea
where we have used 
$\frac{1}{2\pi}\int_0^{\eta'} d\eta \, e^{\pm i(\om-\om')\eta} = \half \d(\om - \om')) \pm \frac{1}{2\pi}\,P\left(\tfrac{1}{\om-\om'}\right)$, and  have dropped the Principal Value integral as the upper limit goes to $\eta'\to\infty$.

We note two features here. First, to $\mO(\b_0)$ we have 
$\sqrt{\tfrac{1\mp \b_0}{1\pm \b_0}} \approx 1\mp \b_0$, so that the frequencies $\om (1\mp \b_0)$ that appear in \Eq{A:s:first:approx} are just the Doppler shifted frequencies we'd expect to  see when Irwin is traveling at nearly constant velocity $\b_0$. Secondly, since we have taken $\om, \Om >0$, and $ (1\mp \b_0)>0$, the frequencies $\Om + \om\,(1\pm \om)>0$ that appear in both delta functions for $\beta^*_{\om \Om}$ are positive, and hence the delta functions are zero. This latter point says that in the limit of $\b_0\ll 1$, Bob's positive frequency plane wave appears also as a positive frequency plane wave to Irwin.

%===================================================================
\subsubsection{\black{The $\mathcal{O}(\beta_0)$ contribution to \Eq{A:s}}}
%===================================================================
\black{
The  $\mathcal{O}(\beta_0)$ contribution to \Eq{A:s} is given by 
\bea{A:s:1}
\mA_s^{(1)} &=& 
\frac{1}{2 \pi a_0}\,
 \left( \frac{-i s \om \beta_0}{a_0}\right)\,
\int_{-\infty}^{\infty} d\eta\, e^{\tfrac{i \Om'_s}{a_0}\,\eta}\,
\ln(1+e^{-2 \eta}), \no
&=& 
\frac{1}{2 \pi a_0}\,
 \left( \frac{-i s \om \beta_0}{a_0}\right)\, 
 \int_0^\infty dy\, y^{\nu-1} \ln(1+y^2), \no
 &=& 
 \frac{1}{2 \pi a_0}\,
 \left( \frac{i s \om \beta_0}{a_0}\right)\, 
\frac{\pi}{\left(\frac{\Om'_s}{a_0}\right)\,\sinh\left(\frac{\Om'_s}{a_0} \right)}
\eea
where we have defined $\Om'_s \defn \Om + s\, \om\,(1- \b_0)$, $y = e^{-\eta}$, and  $\nu = \tfrac{-i \Om'_s}{a_0}$. 
The integral in the second line above was performed using 
$\int_0^\infty dx\, x^{\mu-1}\, \ln(1+x) = \tfrac{\pi}{\mu\, \sin(\mu x)}$
(see integral (4.293.2) p558 of Gradshteyn and Ryzhik\cite{Gradshteyn_Ryzhik:1965}), 
%\tit{Table of Integrals, Series and Products\cite{Gradshteyn_Ryzhik:1965}}), 
making the change of variable $x=y^2$, and using the identity $\sin(-i w) = -i \sinh(w)$.
The point here is that the integral is finite, and the amplitude is proportional to $\beta_0$, so that
$\mA_s^{(1)}\overset{\beta_0\to 0}{\longrightarrow} 0$. Thus, only 
$\mA_s^{(0)}$  in \Eq{A:s:first:approx} remains in this limit, with associated $\beta^*_{\om \Om}= 0$, corresponding to a zero temperature.
} % end red
\subsection{Limit $\boldsymbol{\b_0=1-\eps,\; \eps\ll 1}$}\label{subsec:I2R:Temp}
%==========================================================
In this section we consider the opposite limit of near constant acceleration, i.e. 
$\b_0~=~1~-~\eps,\; \eps\ll~1$.
In this limit we have
\bea{U:beta0:approx:1}
a_0\,U &=& -e^{-\eta} + \frac{\eps}{4}\, (e^{\eta} + \tfrac{1}{3}\,e^{\black{-3\,\eta}}) + \mO(\eps^2), \label{U:beta0:approx:1:line1} \\
\Rightarrow e^{i \,s\, \om\, U} &=& 
e^{- i \,s\, \tfrac{\om}{a_0}\, e^{-\eta}}\,e^{i \,s\, \om\,\frac{\eps}{4}\, (e^{\eta} + \frac{1}{3}\,e^{-3\eta})  }, \no
&\approx& e^{- i \,s\, \tfrac{\om}{a_0}\, e^{-\eta}}\,
\big[1 + i \,s\, \om\,\frac{\eps}{4}\, (e^{\eta} + \tfrac{1}{3}\,e^{-3\eta}) \big], \label{U:beta0:approx:1:line2} 
\eea
where we have used $e^{i\, \eps\, b}\approx 1 + i\, \eps\, b  + \mO(\black{\eps^2})$.
The first term in \Eq{U:beta0:approx:1:line2} is the standard Rindler term for constant acceleration, 
$\b_0\, a_0\overset{\b_0\to 1}{\longrightarrow} a_0$ leading to the standard Unruh temperature $T_U$, while the terms proportional to $\eps$ are correction terms. Letting $y= e^{-\eta}$ and performing the standard Rindler change of variables we obtain 
\bea{A:s:beta:approx:1}
\mA_s & \defn& \frac{1}{2 \pi a_0} \,
\Big[ 
\mA_s^{(\nu_0 = 0)} +
 \frac{i\,s\,\om'}{4}\left(\mA_s^{(\nu_0 = -1)} + \tfrac{1}{3}\,\mA_s^{(\nu_0 = 3)} \right) 
\Big], \\
\trm{where}\quad \mA_s^{(\nu_0)} &=& 
\int_{y=0}^\infty dy\, y^{(-i\,\Om'+ \nu_0)-1}\, e^{-i (s\,\om')\,y} = 
\frac{1}{(i\,\om')^{\nu_0}} \, \G\left(-i\,\Om'+ \nu_0\right)\, e^{i\,\Om'\,\ln \om'},
\eea
where we have defined $\Om'\defn\tfrac{\Om}{a_0}$ and $\om'\defn\tfrac{\om}{a_0}$, and we
have repeatedly use the Gamma function integral formula \Eq{Keifer:p50} with $\nu = -i\Om' \to  -i\Om' + \nu_0$,
with $\nu_0 = 0$ (Rindler case), and $\nu_0\in\{-1, 3\}$ for the correction terms.

If we now consider $\beta^*_{\om \Om}\propto \mA_{s=1}$, we obtain with a little bit of straightforward algebra,
\be{mA:s1:beta0:approx:1:I2R}
\hspace{-0.5in}
|\beta^*_{\om \Om}|^2 \propto 
\frac{2 \pi}{\Om'}\,\frac{1}{(e^{2\,\pi\,\Om'}-1)}\,
\left[
1 - \frac{\eps}{4}
\left(
\frac{\om'^2}{z-1} + 
\frac{1}{3}\,\frac{z\,(z+1)\,(z+2)}{\om'^2}
\right)
\right], \quad z = -i\,\Om' = -i\, \tfrac{\Om}{a_0},
\ee
where we have factored out the Rindler contribution
$|\G(-i\, \tfrac{\Om}{a_0})|^2~=~\tfrac{2\,\pi}{(\Om/a_0)\,\sinh(\pi \Om/a_0)}$, 
and have made repeated use of the Gamma function identity $\G(1+z) = z\,\G(z)$ to obtain the form of the correction terms in \Eq{mA:s1:beta0:approx:1:I2R}. We interpret \Eq{mA:s1:beta0:approx:1:I2R}
as a (modified) Rindler Planck thermal spectrum, essentially arising from the $1$  in $\b_0 = 1-\eps$, and the remainder  as small $\eps$-corrections to the near thermal spectrum, with the Unruh temperature
$k_b\,T_U \overset{\b_0\lesssim 1}{=} \hbar\,a_0/(2\,\pi\,c)$.

%==================================================================
\subsection{\black{An} argument for a proposed \black{natural adiabatic effective $\mathbf{I2R}$ temperature candidate} given by
\black{$\boldsymbol{T_{I2R} = \hbar\,a(t)/(2\,\pi\,c)}$}}\label{subsec:I2R:Temp:Conjecture}
%==================================================================
From the above two limits, we observe that Irwin obtains naturally, for the negative frequency content 
$|\beta_{\om \Om}|^2$ of Bob's purely positive frequency plane wave $e^{-i\,\om\,U}$,  
(i) a zero temperature for small beginning/ending terminal velocities $\b_0\to 0$, and 
(ii) the Planck spectrum with Unruh temperature $k_b\,T_U = \hbar\,a_0/(2\,\pi\,c)$ in the opposite limit 
$\b_0\to 1$.

Physically, we expect that during the predominantly constant velocity portion of the journey with velocity profile $\beta(t) = \b_0\,\tanh(a_0\,t) \approx \pm\,\b_0$ for $|t| \gg \tfrac{1}{a_0}$, that there should be no excitation of particles in the Minkowski vacuum. Therefore, we expect that the perception of particles in the Minkowski vacuum by the traveling observer should appear during the period of acceleration.
In addition, we expect this period of time should be related to the duration when the traveler is 
within the RRW, and is temporarily causally disconnected from receiving signals (at traveling at $45^\circ$ in Minkowski spacetime $(T,X)$) from the LRW.
This is the time $\t^*(\b_0)$ we plotted in \Fig{fig:T:t:Irwin:in:RRW} at which the acceleration 
(at the co-moving observer's proper time $t=0$) falls to half its peak value, 
i.e. $a(\t^*) = \half\,\b_0\,a_0$.

As such, we propose that a   
\black{natural adiabatic I2R effective temperature candidate be given by} 
%I2R \black{effective} temperature 
%can be given by
\be{TI2R:defn}  
T_{I2R} \defn \frac{\hbar\, a(t)}{2\,\pi\,c}\approx  \frac{\hbar\, \b_0\,a_0}{2\,\pi\,c}
\;\; \trm{for}\;\; t\in [-\t^*(\b_0),\, \t^*(\b_0)].
\ee
We note that \Eq{TI2R:defn} reproduces the zero temperature and Unruh temperature in the limits
$\b_0\to 0$ and $\b_0\to 1$, respectively. \black{Heuristically}, we also propose that this temperature is perceived by the co-moving observer Irwin essentially only while Irwin's orbit is within the RRW $t\in [-\t^*(\b_0),\, \t^*(\b_0)]$ when he is causally disconnected from receiving any signals from ``mirror" observers in the LRW. As $\b_0$ approach unity, Irwin's orbit is more and more confined to the RRW, as shown in  \Fig{fig:oribits:lines:of:simultaneity:2}(right) and 
\Fig{fig:T:t:Irwin:in:RRW} and so his temperature becomes more and more like the uniformly accelerate Rob's Unruh temperature, as the duration of acceleration becomes infinite. 

\black{
Strictly speaking, a temperature can only be defined for static or stationary spacetime metrics where a well defined Killing vector can be defined (e.g. $\pd_T$ for Minkowski spacetime, $\pd_t$ for Rindler spacetime) so that positive $f_k$ and negative $f_k^*$ frequency modes of particles can be well defined, 
(i.e. $\pd_t\, f_k = -i\,\om\,f_k$ for $1+1$ dimensional positive frequency modes  $\om=|k|>0$ with wavevector $k$; for Rindler\cite{Carroll:2004} $f_k(t,\xi)= (4\pi\om)^{-1/2} \,e^{i\,(k\,\xi - \om\,t)}$).
For metrics which depend explicitly on $t$, (such as the I2R discussed here
 through the non-stationary acceleration $a(t)$), the notion of particles is not, in general, a well defined concept (see Chapter 9 of Carroll\cite{Carroll:2004} for a more complete discussion involving general curved spacetimes). 
 It is more meaningful to consider instead the response of a (Unruh-DeWitt) detector along a trajectory undergoing arbitrary motion.
 In their classic work \tit{Quantum Fields in Curved Space}\cite{Birrell_Davies:1982}, Birrell and Davies  show that for \tit{any} motion \tit{not} at constant velocity, the (arbitrary) accelerated observer will perceive 
some response to a local particle detector. In fact the local detector will both be excited, and also \tit{emit} a particle (see Unruh and Wald\cite{Unruh_Wald:1982}). This counterintuitive phenomena arises   
because the particle detector is ultimately excited by the force providing the acceleration (Irwin's rockets). 
% %
% Such intuition is also born out by the many examples examined in 
%the classic work    \tit{Quantum Fields in Curved Space}\cite{Birrell_Davies:1982} by Birrell and Davies, who show that for \tit{any} motion \tit{not} at constant velocity, an (arbitrary) accelerated observer will perceive some response to a local particle detector.
In the gravitational case, the Hawking radiation is ultimately produced by the acceleration $=$ surface gravity of the BH, which shrinks in radius as it looses mass resulting from the production of the Hawking radiation.
In the cosmological case, an expanding universe ($ds^2 = dt^2 - a^2(t)\,dx^2$, with
expansion parameter $a^2(t)=A + B\tanh(\alpha\,t)$ yielding asymptotic beginning and ending flat regions), will create particles\cite{Birrell_Davies:1982} primarily peaking around the rapid transition phase (peak acceleration) about $t=0$. This feature is essentially analogous to the cause for particle creation in our current universe during the period of cosmological inflation (arising from the Higgs field achieving  a non-zero expectation value as it decays into a new ground state). Additionally, it is also well known that in the sudden approximation, a quantum harmonic oscillator abruptly changing frequencies from $\om_i$ to $\om_f$ will generate a single-mode squeezed (final vacuum) state from an initial (empty) vacuum state, i.e. there will be non-zero particles in the new vacuum state (see Problem 2.10 in Agarwal\cite{Agarwal:2013}).
} % end of red
 
\black{ 
Additionally, a crucial consideration for the definition of the Unruh temperature for the Rindler metric is that all observer trajectories are confined for all time to the causally disconnect regions of the RRW, and the (``mirror" observer) LLW. As briefly discussed earlier, quantum mechanically this means that the quantum state of observer in the RRW is formed by tracing out over the modes of the LRW, leading to a RRW mixed state that can be interpreted as an exact thermal state with a Planckian particle spectrum (again, see Chapter 9 of Carroll\cite{Carroll:2004} for further details). As discussed earlier,  the trajectories of the I2R observer are confined  to the ``extended wedges" bounded by the
asymptotes $X/T = \pm\beta_0$ which encompass the RRW/LRW for $0<\beta_0<1$. Thus, any associated ``temperature" cannot be strictly thermal (as is the Unruh temperature for Rindler uniformly accelerated trajectories) since these I2R regions are \tit{not} 
causally disconnected \tit{for all time}.
} % end of red  

\black{The effective Unruh temperature we propose is only approximate in the sense that it is more appropriate the more the I2R acceleration $a(t)$ has a wider, flatter plateau for some period of time (as is especially seen for large values of $\beta_0$ less than unity), or is adiabatically changing  such that $\dot{a}(t)/a^2(t) \ll 1$. 
Inserting back factors of $c$, $\hbar$ and $k_b$, and noting that both
$\dot{a}(t)/a(t)$ and $a(t)/c$ have units of frequency (sec$^{-1}$), 
this latter quantity can be interpreted as the  relative  change of Unruh thermal energy due to its acceleration, to the instantaneous Unruh thermal energy defined as proportional to its instantaneous acceleration $a(t)$, i.e.  $\frac{\hbar\, \dot{a(t)}/a(t)}{\hbar a(t)/c} = \frac{k_b\,\Delta T_U(t)}{k_b\,T_U(t)}$.
Analogous results were found by Obaida and Milgrom\cite{Obadia_Milgrom:2007} 
in their paper ``\tit{Unruh effect for general trajecctories}" in which 
they explored the applicability of the quasistationary approximation, whereby the response function of an accelerated detector is approximated by its value for stationary trajectory with the instantaneous values of the invariant parameter of the acceleration, torsion and hypertorsion (the latter two of which are scalar functions of the 4-acceleration $a^\mu$ (see their Eqs.(10-12)). Their work found numerical validity of such an approach in the low energy regime where $E/a<1$, and 
$\dot{a}(t)/a^2(t) \ll 1$ (with $c=1$).
}% end of red

\black{
Additionally, Barbado and Visser\cite{Barbado_Visser:2012} performed an complementary analysis for observers with arbitrary accelerations. The authors computed the response function and click rate of an Unruh-DeWitt detector in terms of an adiabatic expansion  in the derivatives of the acceleration $a(t)$.  The $k$-th order coefficients of the adiabatic expansion involve
ratios of the form $\frac{a^{(k)}(t)}{a(t)^{k+1}}$ for $k\ge 1$, where 
$a^{(k)}(t)\defn \frac{d^ka(t)}{dt^k}$ is the $k$-th derivative of $a(t)$.
All orders of the detector response function contain a multiplicative factor of the instantaneous acceleration $a(t)$.
To lowest ($0$-th) order, the detector response function was simply proportional to the instantaneous acceleration $a(t)$, and to $1$st order, agrees with the previous result containing the additional multiplicative ratio 
$\dot{a}(t)/a^2(t)$ of Obaida and Milgrom\cite{Obadia_Milgrom:2007}. 
Barbado and Visser interpret their adiabatic expansion, acceleration dependent
response function as a ``\tit{quantity capturing the physical intuition that, during
periods in which an acceleration is present, the detector
would be continuously encountering new particles, producing some sort of cumulative effect}"\cite{Barbado_Visser:2012}. 
Thus, our 
natural adiabatic effective $I2R$ temperature candidate given by
$T_{I2R} = \hbar\,a(t)/(2\,\pi\,c)$ is in line with Barbado and Visser's zeroth-order result if we consider  accelerations with are adiabatically varying.
Further investigations for an adiabatic Unruh temperature arising from  I2R, and arbitrary, accelerations in flat spacetime  are explored by the author in \cite{Alsing:WKB:paper:arxiv}.
} % end of red

\black{
Heuristically, we expect that particle creation for the I2R trajectories will occur in region of 
effective non-zero acceleration, which predominantly occurs when the trajectory is within
the RRW (since the I2R trajectory asymptotes rapidly to $X/T = \pm \beta_0$ outside the RRW).
By adjusting the two I2R parameters $(\b_0, a_0)$, one can arrange for a nearly constant acceleration of $a(t) \approx \b_0\,a_0$ over the period $|t|\le \t^*(\b_0)$, the relevant integrals for $\beta_{\om \Om}$ are essentially the Rindler integral in 
\Eq{alpha:beta} with (i) $a_0\to \b_0\,a_0$, and (ii) the infinite limits of integration replaced by 
$\pm \infty \to \pm \t^*(\b_0)$. Such finite-limit integrals can be \tit{formally} written in terms of the difference of two incomplete gamma functions $\g(\nu,\t)$ (with complex arguments) 
where  $\G(\nu,\t) \defn \int_{0}^{\t} dz\, z^{\nu-1}\,e^{-z}$ 
(vs. $\G(\nu) = \int_{0}^{\infty} dz\, z^{\nu-1}\,e^{-z}$). 
However, $\G(\nu,\t)$ does not have nice analytic forms/properties, as does its parent $\G(\nu)$, and so it is only of possible use for numerical investigation.
On the other hand, one could also choose $a_0\gg 1$, but $\b_0 \ll 1$ such that $\b_0\,a_0\to$ finite.
Then the $\sinh(a_0 t)$ is sharply spiked about $t=0$ and one can approximate the acceleration as 
 $a(t)\approx \b_0\,a_0\,\d(t)$ (or equivalently, a rectangular function of constant height 
$\b_0\,a_0$ over the brief period $t\in [-\t^*(\b_0),\, \t^*(\b_0)]$). This is the ``immediate turnaround case, for which there would be a ``delta-function" burst of particles perceived by Irwin at temperature 
$T_{I2R}~=~\hbar (\b_0\, a_0)/(2\,\pi\,c)$ for an instant, so in all practical effect, Irwin would  register a zero temperature.
} % end of red

\black{
We note again that \Eq{TI2R:defn} reproduces the zero temperature 
and Unruh temperature in the limits
$\b_0\to 0$ and $\b_0\to 1$, respectively. We also propose that this 
adiabatic temperature candidate
is perceived by the co-moving observer Irwin essentially only while Irwin's orbit is within the RRW $t\in [-\t^*(\b_0),\, \t^*(\b_0)]$ when he is causally disconnected from receiving any signals from observers in the LRW. As $\b_0$ approach unity, Irwin's orbit is more and more confined to the RRW, as shown in  \Fig{fig:oribits:lines:of:simultaneity:2}(right) and 
\Fig{fig:T:t:Irwin:in:RRW} and so his temperature becomes more and more like the uniformly accelerate Rob's Unruh temperature, as the duration of acceleration becomes infinite. 
} % end of red

\section{Conclusion}\label{sec:Conclusion}
%==================================================================
In this work we have examined motion in flat (Minkowski) spacetime where the accelerated observer Irwin with coordinates $(t, x)$ has a velocity profile $\beta(t)$ (with $c=1$) with respect to an inertial observer Bob with coordinates  $(T,X)$. In general, if we require that Irwin's starts his trip at some  large positive distance $X_0>0$, travels towards the origin $X=0$ with negative (varying) velocity, and returns to his starting point  with positive velocity, then $\beta(t)$  must be an odd function under $t\to -t$. Since we want the magnitude of the velocity to be constant at the beginning and end of the trip, one expects
the velocity profile to have the form $\beta(t) = \tfrac{f(t)}{g(t)} $ where $f(t), g(t)$ are odd/even under time inversion, and we physically require that $|\beta(t)|\le 1$. In this present work, we considered 
\black{ as a ``natural choice"}  $\beta(t) = \b_0\,\tanh(a_0\,t)$, since this has the inertial and Rindler profiles as $\b_0\to 0$ and $\b_0\to 1$, respectively, and the resulting integrals for $T(t) = \int dt\, \g(t)$ and $X(t) = \int dt\, \beta(t)\,\g(t)$ (where $\g(t) = (1-\beta^2(t))^{-1/2}$)
yield closed form solutions. 

However, this is not the only choice. If one allowed Irwin to leave Earth located at some distance $-X_0$ at some large negative time ($T_0\to -\infty$) with zero velocity (matching the stationary Bob), and travels  to Alice at Proxima Centauri at $X_0$ at some large positive time ($T_0\to \infty$), then one could, for example, choose the purely positive velocity profile $\beta(t) = \b_0/\cosh(a_0\,t)$ which yields
\bea{pos:vel:profile:T:X}
T(t) &=& \frac{1}{a_0}\,\sinh^{-1}(\g_0\,\sinh(a_0\,t)),\no
 X(t) &=& \frac{k}{a_0}\, F(\theta, k),\quad  F(\theta,k) \defn \int_0^\theta\, \frac{dx}{\sqrt{1-k^2\,\sin^2 x}},\quad x=a_0\,t, \no
 k &=& \frac{\b_0\,\g_0}{\sqrt{1+\b^2_0\,\g^2_0}}, \quad 
 \theta(x) = 
 \sin^{-1}
 \left( 
 \frac{\sqrt{1+\b^2_0\,\g^2_0}\,\tanh(x)}
 {\sqrt{1+\b^2_0\,\g^2_0\,\tanh(x)}}
 \right),
\eea
where $F(\theta, k)$ is the incomplete Elliptic integral of the first kind. (Note the inverse placement of $\g_0$ inside the arcsinh, and lack of it outside, compared to the I2R coordinate $T_{I2R}(t)$). Such a transformation is a more ``realistic" version of the Twin ``Paradox," since Irwin matches Bob's and Alice's velocities at the beginning and end of his journey, respectively, with peak acceleration in the middle at $X(t=0)=0$. More importantly, this trajectory more closely resembles the asymptotically flat cosmological expansion scenario considered by Birrell and Davies\cite{Birrell_Davies:1982} (discussed above, with a $a^2(t) = A + B\,\tanh(\alpha t)$ expansion factor profile) in which particle creation occurs (and one can also analytically compute 
 $\alpha_{\om \Om}$ and $\beta_{\om \Om}$). Thus, one expects to associate a slowly time varying  temperature to this transition if the expansion is carried out adiabatically, although the concept of a ``particle" is strictly not well defined in the non-asymptotic acceleration regions (see Birrell and Davies\cite{Birrell_Davies:1982}, and Chapter 9 of Carroll \cite{Carroll:2004}).

For the I2R coordinates studied here, we can adjust the parameter $(\b_0, a_0)$ so that the acceleration region is either abrupt, or gradual, with varying magnitude of the peak acceleration at $a(0) = \b_0\,a_0$. Therefore, it is reasonable to propose 
\black{a natural adiabatic effective $I2R$ temperature candidate}
%an associated time varying I2R temperature 
$T_{I2R}(t) \defn \frac{\hbar\, a(t)}{2\,\pi\,c}$, for fixed values of $(\b_0, a_0)$, at least during effective period of acceleration $t\in [-\t^*(\b_0),\, \t^*(\b_0)]$ (when Irwin is within the RRW, as discussed in \Eq{TI2R:defn}). 

\black{As discussed above, the
\black{effective adiabatic temperature candidate} 
%effective Unruh temperature 
we propose is only approximate in the sense that it is more appropriate the more the I2R acceleration $a(t)$ has a wider, flatter plateau for some period of time (as is especially seen for large values of $\beta_0$ less than unity), or is adiabatically changing  
such that $\dot{a}(t)/a^2(t) \ll 1$. 
\black{
Such a natural adiabatic temperature candidate proportional to the observer's instantaneous acceleration $a(t)$ has support from the previous research of Obaida and Milgrom\cite{Obadia_Milgrom:2007}  and Barbado and Visser\cite{Barbado_Visser:2012} in their studies of the response of an Unruh-DeWitt detector for observers undergoing arbitrary acceleration in flat spacetime.
}% end red
%
%Analogous results were found by Obaida and Milgrom\cite{Obadia_Milgrom:2007} 
%%in their paper ``\tit{Unruh effect for general trajecctories}" 
%in which they explored the applicability of the quasistationary approximation, whereby the response function of an accelerated detector is approximated by its value for stationary trajectory with the instantaneous values of the invariant parameter of the acceleration, torsion and hypertorsion (the latter two of which are scalar functions of the 4-acceleration $a^\mu$ (see their Eqs.(10-12)). Their work found numerical validity of such an approach in the low energy regime where $E/a<1$, and 
%$\dot{a}(t)/a^2(t) \ll 1$. 
%\black{Similar results were found by Barbado and Visser\cite{Barbado_Visser:2012} for an Unruh-DeWitt detector undergoing arbitrary acceleration when the response function was adiabatically expanded in terms of higher order time derivatives of the acceleration. The $0$th order response function was simply proportional to the instantaneous acceleration $a(t)$, and the $1$st order response function was proportional to  $\dot{a}(t)/a^2(t)$, agreeing with the results of 
%Obaida and Milgrom\cite{Obadia_Milgrom:2007}.
%}% end red
}% end of \black{}

\black{
Our work is clearly not the only one that has considered exploration of the Unruh effect for non-stationary trajectories.
Early relevant works in this area include those of 
Letaw and Pfautsch (1980)\cite{Letaw_Pfautsch:1980}, Letaw (1981)\cite{Letaw:1981}, and Bell and Leinass (1983)\cite{Bell_Leinaas:1983}, as well as more recent works of Louko and Satz (2006)\cite{Louko_Satz:2006}, 
\black{the extensive Unruh effect review article by Crispino \tit{et al.}\cite{Crispino:2008}},
and the above mentioned Obadia and Milgrom (2007)\cite{Obadia_Milgrom:2007} \black{and Barbado and Visser\cite{Barbado_Visser:2012}}. We feel that this present work complements and adds this body of investigations by presenting a simple, tractable, and physically motivated two-parameter 
$(a_0,\beta_0)$ generalization of well-studied one-parameter ($a_0$) uniform acceleration Rindler transformation.
} % end of \red
%\bk

\black{
Another area of interest for this work is  its extension to non-stationary metrics in curved spacetime, and an analogous formula for the Hawking temperature. The recent work of Hari and Kothawala (2021)\cite{Hari_Kothawala:2021} 
%(K. Hari and D. Kothawala, Phys. Rev. D \tbf{104}, 064032 (2021)) 
 examined the effect of tidal curvature in general curved spacetime on the dynamics of accelerated probes by developing (to lowest order) a coordinate transformation between an inertial observer on a geodesic path, and that of an accelerated observer, by means of an covariant power series expansion in Riemann normal coordinates. Hari and Kothawala developed a modified Unruh temperature in which 
$a_0\to \sqrt{a^2_0\,- \mathcal{E}_n}$, where 
$\mathcal{E}_n = R_{abcd}\,\epsilon^{ab}\,\epsilon^{cd}$ is a purely tidal component of curvature (proportional to the extrinsic curvature) with $\epsilon^{ab}$ the bi-normal to the instantaneous (osculating) plane of (hyperbolic) motion. Interestingly, the formula they develop for the transformation between the proper time of the inertial $\tau_{geo}$, geodesic observer and the accelerated observer $\tau_{accel}$ 
is given by
\be{Hari:Kothawala:eqn}
\tau_{geo} = \tfrac{2}{\sqrt{- \mathcal{E}_n}} \, \sinh^{-1}\,
\Big[
\sqrt{\tfrac{- \mathcal{E}_n}{a_0^2 - \mathcal{E}_n}} \,
\sinh\big[\tfrac{1}{2}\, \sqrt{a_0^2 - \mathcal{E}_n}\big] \, \tau_{accel}
\Big].
\ee
The above equation  was derived by inferring that a laborious 10th order Taylor series expansion 
(in $\tau_{accel}$) of a formula relating  \black{the} accelerated trajectory to the inertial trajectory in Riemann normal coordinates, were the first 10 terms of the Taylor series expansion of \Eq{Hari:Kothawala:eqn}.
Surprisingly, \Eq{Hari:Kothawala:eqn} is  \tit{exactly} the form of the  formula 
for the temporal I2R coordinate $T$ in \Eq{T:X:I2R}, with the identifications
$T\leftrightarrow \tau_{geo}/2$, 
$t\leftrightarrow \tau_{accel}/2$, 
and the  I2R substitutions
$a_0\to \sqrt{a_0^2 - \mathcal{E}_n}$,
$\gamma_0 \to \sqrt{a_0^2 - \mathcal{E}_n}/\sqrt{- \mathcal{E}_n}$, and
$\beta_0 \to a_0/\sqrt{a_0^2 - \mathcal{E}_n} < 1$ for $- \mathcal{E}_n > 0$.
%developed  by the authors (of this manuscript) in a recent 2026 paper (in review at AJP) between a flat spacetime inertial observer and an accelerated observer with velocity profile $\beta(t) = \beta_0\,\tanh(a_0\,t) = \tanh(\chi(t))$ with  $|\beta_0|<1$. 
%Thus, in flat spacetime with $\mathcal{E}_n\to 0$ we have \Eq{Hari:Kothawala:eqn}$\to$\Eq{T:X:I2R}.
Note that as $\mathcal{E}_n\to 0$, the above association yields $\beta_0\to 1$ and $a_0\to \sqrt{a_0^2 - \mathcal{E}_n}\to a_0$ which is our I2R constant acceleration Rindler limit. 
On the other had, as $a_0\to 0$,  the association yields $\gamma_0\to1$ and $\beta_0\to 0$ which is our I2R zero acceleration limit.
Most likely, this correspondence arises due to the instantaneous Lorentz transformation from the inertial geodesic observer and the accelerated observer using the instantaneous velocity profile $\beta(t)$, though role of tidal term  
$ \mathcal{E}_n$, and the reason it acts as a negative acceleration to $a_0$ would be of interest for further investigation. We feel the intriguing result of Hari and Kothwala lend further relevance to our I2R coordinate transformation.
}% end \black{}

In future work, we will consider the effect of this proposed adiabatic temperature \black{candidate} on black hole evaporation when considered as a tunneling process\cite{Srinivasan:1999, Wilczek:2000, Srinivasan:2001, deGill:2010,Alsing:WKB:paper:arxiv}.

%================================
\clearpage
\newpage
%================================
\appendix*
\section{Codes to numerically produce acceleration profiles, orbits  and simultaneity plots}\label{app:Mathematica:codes}
\begin{figure}[h!]
\begin{center}
\begin{tabular}{ccc}
\hspace{-0.5in}
\includegraphics[width=3.5in,height=4.5in]{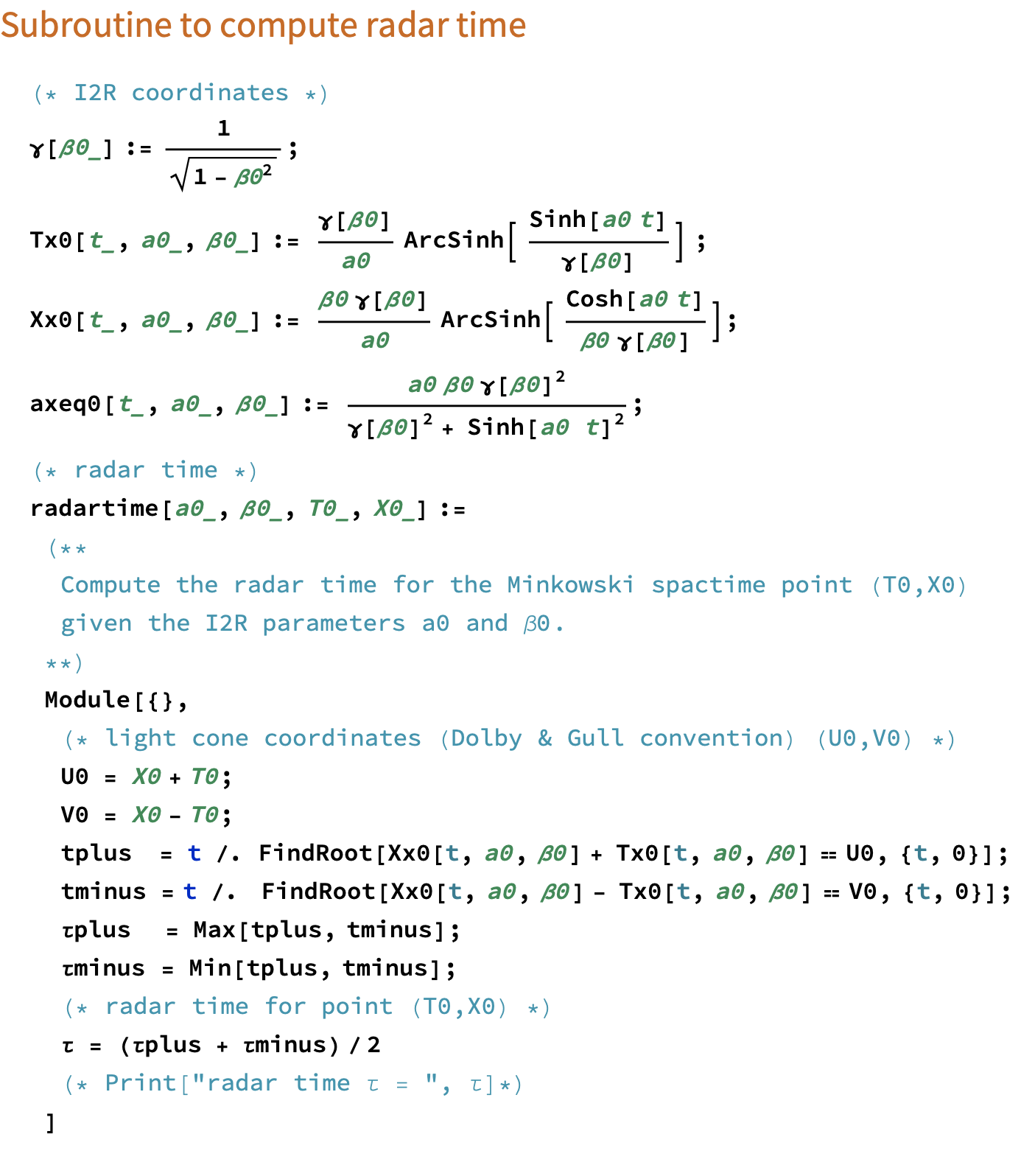} &
{\hspace{0.35in}} &
\hspace{-0.5in}
\includegraphics[width=4.0in,height=5.5in]{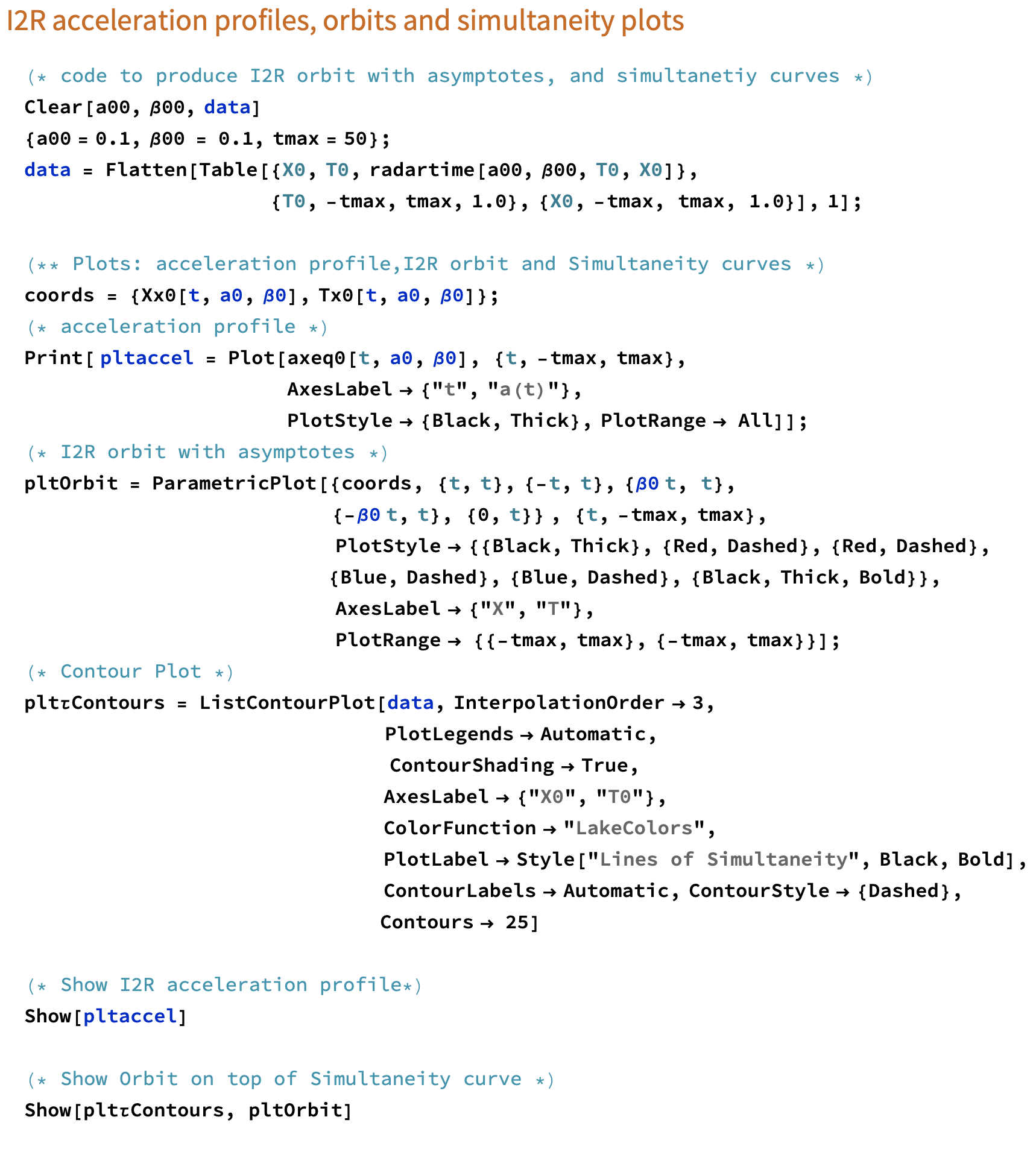}
\end{tabular}
\end{center}
\caption{\tit{Mathematica} codes to 
(left) compute radar time, and 
(right) compute/plot simultaneity curves with overlay of I2R orbit with asymptotes, and acceleration profiles,
used to create the plots in \Fig{fig:oribits:lines:of:simultaneity:1} and \Fig{fig:oribits:lines:of:simultaneity:2}, and in
\Fig{fig:acceleration:profile:1} and \Fig{fig:acceleration:profile:2}.
}\label{fig:Mathematica:codes}
\end{figure}

The following \tit{Mathematica} codes in \Fig{fig:Mathematica:codes} were used to  
(left) compute radar time, and 
(right) compute/plot simultaneity curves with overlay of the I2R orbit with asymptotes, and acceleration profiles,
for the plots  in \Fig{fig:oribits:lines:of:simultaneity:1} and \Fig{fig:oribits:lines:of:simultaneity:2},
following the analytical procedure of Dolby and Gull\cite{Dolby_Gull:2001},
as well as the acceleration profiles in
\Fig{fig:acceleration:profile:1} and \Fig{fig:acceleration:profile:2}.
%================================
%============================
% figures with tabular array
%============================
\clearpage
\newpage
%================================
\begin{acknowledgments}
\vspace{-0.2in}
\black{
The author would like to thank K. Hari and  D. Kothawala for informative discussions on their work.
The author would also like to thank two anonymous referees for their detailed and insightful comments and suggestions that  have helped to improve this work.} % end red

The author has no competing interests for this work. 

Data Statement: The \tit{Mathematica} codes that were used to construct 
\Fig{fig:oribits:lines:of:simultaneity:1} - \Fig{fig:acceleration:profile:2}
are openly available at 
{\verb+https://doi.org/10.7910/DVN/GELDOY+}.
%{\verb+https://doi.org/10.7910/<to be filled in if published>.+}
%} % end red
%{\verb+https://dataverse.harvard.edu/dataverse/alsingpm_research+}. 
\end{acknowledgments}
%================================

%================================
% Create the reference section using BibTeX:
%\nocite{apsrev41Control}
%\bibliographystyle{ieeetr}
%\bibliographystyle{apsrev4-1}
%%%\bibliography{alsing_composite_bibfile}
%================================
% paste in the .bbl file here
%================================
%merlin.mbs apsrev4-1.bst 2010-07-25 4.21a (PWD, AO, DPC) hacked
%Control: key (0)
%Control: author (8) initials jnrlst
%Control: editor formatted (1) identically to author
%Control: production of article title (-1) disabled
%Control: page (0) single
%Control: year (1) truncated
%Control: production of eprint (0) enabled
\providecommand{\noopsort}[1]{}\providecommand{\singleletter}[1]{#1}%
%
%====================================

%====================================
\end{document}